\documentclass[11pt,a4paper]{article}
\usepackage{jheppub,xcolor}
\usepackage[utf8]{inputenc}
\graphicspath{{./figs/}}

\usepackage{amssymb}
\usepackage{dcolumn}	
\usepackage{bm}			
\usepackage{bbm} 		
\usepackage{multirow}
\usepackage{slashed}
\usepackage{blkarray}
\usepackage{booktabs}
\usepackage{makecell}
\usepackage{longtable}
\usepackage{placeins}
\usepackage{graphicx}
\usepackage{amsmath}
\usepackage{float}
\usepackage[font=small,labelfont=bf]{caption}
\usepackage{etoolbox}
\usepackage{subfig}
\usepackage[countmax]{subfloat}

\definecolor{Red}{rgb}{1.,0.,0.}

\newcolumntype{?}[1]{!{\vrule width #1}}

\newcommand{\bc}{\begin{center}}
\newcommand{\ec}{\end{center}}
\newcommand{\ba}{\begin{array}}
\newcommand{\ea}{\end{array}}

\begin{document}

\title{Search prospect for electrophilic bosons at the $\mu$TRISTAN}

\author[a]{Amit Adhikary,}
\author[b]{Anindya Datta,} 
\author[a]{Dilip Kumar Ghosh}

\affiliation[a]{School of Physical Sciences, Indian Association for the Cultivation of Science, \\
2A $\&$ 2B, Raja S.C. Mullick Road, Kolkata, West Bengal-700032, India} 

\affiliation[b]{Department of Physics, University of Calcutta, 92 Acharya Prafulla Chandra Road, Kolkata 700009,
India}

\emailAdd{amit93.adhikary@gmail.com}
\emailAdd{adphys@caluniv.ac.in}
\emailAdd{tpdkg@iacs.res.in}

\abstract{Mediators coupled predominantly to the electron remain weakly constrained above the kinematic reach of the $B$~factories, a region that is difficult to access at hadron colliders and only partially covered by existing $e^+e^-$ data. We point out that the asymmetric $\mu^+e^-$ mode of the proposed $\mu$TRISTAN collider, operating at $\sqrt{s}=346$~GeV, is well suited to this window, and quantify its reach for two minimal realisations: an axion-like particle and a $Z'$ gauge boson, treated as effective interactions with the electron current. Both are radiated off the electron line and decay to an electron-positron pair, giving a common four-lepton signature whose Standard Model counterpart is under good control. A resonance search in the electron-positron invariant mass, performed for mediator masses between $5.5$ and $300$~GeV, yields projected $95\%$~CL sensitivities to couplings of order $10^{-5}$ at the light end of this range, improving on existing bounds by more than four orders of magnitude for the pseudoscalar and by up to two orders of magnitude for the vector, and providing the only constraint on the heaviest masses considered.}

\maketitle

\section{Introduction}
\label{sec:intro}

\section{Introduction}
\label{sec:intro}

The discovery of the $125$~GeV Higgs boson at the Large Hadron Collider (LHC)~\cite{ATLAS:2012yve,CMS:2012qbp} completed the particle content of the Standard Model (SM) and confirmed its remarkable predictive power. The framework nevertheless leaves several empirical facts unexplained. It offers no candidate for the dark matter whose presence is inferred from several astrophysical and cosmological observations~\cite{Zwicky:1933gu,Rubin:1970zza,Clowe:2006eq,Planck:2018vyg}. In the lepton flavour sector, the SM does not accommodate the neutrino masses established by various neutrino oscillation experiments~\cite{DayaBay:2012fng,RENO:2012mkc,MINOS:2013xrl}. Furthermore, the CP violating phase present in the CKM matrix of the quark sector is not large enough to explain the observed baryon asymmetry of the Universe~\cite{Sakharov:1967dj,Planck:2018vyg,ParticleDataGroup:2024cfk}; and it provides no dynamical understanding of the smallness of the QCD $\theta$ parameter~\cite{Peccei:1977hh,ParticleDataGroup:2024cfk}. Each of these shortcomings motivates an extension of the SM, and many well-motivated beyond the SM scenarios involve light and weakly coupled states rather than heavier ones. Such states are not necessarily excluded by the null results of high-energy searches, since their production rates are suppressed by small couplings rather than by kinematics, and their identification requires dedicated strategies at both high-intensity and high-precision facilities~\cite{Essig:2013vha,Alekhin:2015byh,Harris:2014hga}. A simple example is a moderately light bosonic state that couples predominantly, or even exclusively, to electrons. An especially interesting mass window extends from $\mathcal{O}(10~\mathrm{GeV})$ to a few hundred GeV. The lower part of this range is probed by $B$-factory experiments, while the upper end is constrained mainly by LEP contact-interaction searches. This leaves a phenomenologically interesting region that is particularly well suited to a collider operating at centre-of-mass energies of a few hundred GeV and with an electron in the initial state, allowing the new boson to be radiated directly from the incoming electron. In the following, we will consider a couple of representative scenarios of this kind.

A prominent example is the axion, originally introduced to relax the strong CP problem through the spontaneous breaking of a global chiral symmetry~\cite{Peccei:1977hh,Weinberg:1977ma,Wilczek:1977pj,Kim:2008hd}, and realised in a variety of ultraviolet completions~\cite{Kim:1979if,Shifman:1979if,Zhitnitsky:1980tq,Dine:1981rt}. The term axion-like particle (ALP) is used more broadly for any pseudo-Nambu-Goldstone boson arising from the breaking of a global symmetry at a scale above the electroweak one, without necessarily addressing the strong CP problem~\cite{DiLuzio:2020wdo,Bharucha:2022lty,Biekotter:2025fll}. In this wider sense, the mass and the couplings of the ALP become independent parameters, which considerably enlarges the accessible phenomenology. ALPs have been studied as dark matter candidates and as mediators to a dark sector~\cite{Dine:1982ah,Preskill:1982cy,Abbott:1982af,Arias:2012az,Jaeckel:2014qea,Co:2019jts,Bharucha:2022lty,Ghosh:2023tyz}, and they leave imprints in cosmology~\cite{Marsh:2015xka,Cadamuro:2010cz,Caloni:2022uya} and in stellar and supernova environments~\cite{Turner:1987by,Raffelt:1996wa,Caputo:2024oqc}. On the experimental side, their effective couplings to SM states have been constrained by a broad programme of collider, beam-dump and flavour-factory measurements~\cite{Bauer:2017ris,OPAL:2002vhf,Jaeckel:2015jla,Mimasu:2014nea,Brivio:2017ije,Bonilla:2022pxu,Mitridate:2023tbj,Dutta:2023abe,Adhikary:2024mzi,CHARM:1985anb,Bjorken:1988as,Blumlein:1990ay,Dobrich:2017gcm,NA64:2020qwq,BaBar:2014zli,Inguglia:2016acz}, and their prospects at future lepton machines have also been examined~\cite{Bao:2022onq,Yue:2021iiu,Tian:2022rsi}. The bulk of this effort, however, has been directed at the ALP couplings to photons and gluons, for which the production mechanisms are most efficient at hadron colliders and in optical or beam-dump experiments. Couplings to charged leptons, and to the electron in particular, are considerably less well covered, and the constraints on them weaken sharply once the ALP mass exceeds the reach of the $B$~factories.

A second, and structurally different, possibility is that the new state is a spin-one boson associated with an additional abelian gauge symmetry. Such $U(1)_X$ extensions arise naturally in constructions addressing the dark matter portal~\cite{Okada:2016tci,Mohapatra:2020bze,Okada:2021nwo}, the origin of neutrino masses~\cite{Ma:2015raa,Okada:2021nwo}, the baryon asymmetry~\cite{Qi:2024pqe} and the reported flavour anomalies~\cite{Bonilla:2017lsq}, and their phenomenology has been reviewed extensively~\cite{Bauer:2018onh,Coloma:2020gfv,Coloma:2022umy}. The resulting $Z'$ boson has been searched for across a wide range of masses and couplings: at the LHC and LEP for TeV- and sub-TeV-scale masses~\cite{CMS:2016cfx,ATLAS:2019erb,Das:2016zue,Accomando:2017qcs,Essig:2009nc}, at LHCb and the $B$~factories in the GeV regime~\cite{LHCb:2017trq,BaBar:2014zli,Inguglia:2016acz}, and in fixed-target, neutrino-scattering, cosmological and astrophysical settings for lighter mediators~\cite{Bross:1989mp,COHERENT:2021xmm,Coloma:2022umy,Escudero:2019gzq,Ghosh:2024cxi,Kolb:1987qy,Croon:2020lrf,Akita:2023iwq,Li:2023vpv}. As in the ALP case, the strongest of these bounds rely either on couplings to quarks, which are absent for a leptophilic mediator, or on the kinematic reach of low-energy machines.

These considerations point to a clear limitation of the existing collider facilities for searches for electrophilic mediators in the intermediate-mass regime. For example, at the LHC, the overwhelmingly hadronic background makes such searches comparatively challenging. On the other hand, conventional $e^+e^-$ colliders provide a much cleaner environment and direct access to electron couplings, but their sensitivity is restricted by their available centre-of-mass energy and luminosity, leaving a significant region above the $B$-factory reach only weakly explored. A recent study~\cite{Adhikary:2026alp} has shown that the Electron--Ion Collider can extend the sensitivity to such electrophilic bosons into this intermediate-mass window, demonstrating the value of having an energetic electron in the initial state. This naturally motivates the study of a dedicated high-energy lepton collider such as $\mu$TRISTAN \cite{Hamada:2022mua}. In its asymmetric $\mu^+e^-$ mode, $\mu$TRISTAN combines a high-energy electron beam with a TeV-scale muon beam, providing a centre-of-mass energy in the few-hundred-GeV range while retaining a clean leptonic initial state. An electrophilic mediator can then be radiated directly from the incoming electron, with substantially reduced hadronic backgrounds and kinematic reach extending well beyond that of the $B$ factories. The $\mu^+e^-$ configuration is therefore particularly well suited to probing weakly coupled electron-specific resonances in the mass range considered here.

In this work, we consider the $\mu$TRISTAN proposal~\cite{Hamada:2022mua}, a facility at J-PARC that uses ultra-cold muon technology to produce a low-emittance $\mu^+$ beam. Two operating modes have been envisaged: a $\mu^+\mu^+$ mode with two $1$~TeV beams, corresponding to $\sqrt{s}=2$~TeV, and an asymmetric $\mu^+e^-$ mode in which a $1$~TeV muon beam collides with a $30$~GeV electron beam, corresponding to $\sqrt{s}=346$~GeV with an integrated luminosity of up to $1000~{\rm fb}^{-1}$. It is the latter configuration that we exploit here. The $\mu^+\mu^+$ mode has no electron in the initial state, so that an electrophilic mediator cannot be radiated directly off a beam particle and could only be produced from a photon-initiated electron pair, at the price of additional powers of the electromagnetic coupling. Its higher centre-of-mass energy would nevertheless extend the accessible mass range well beyond the one considered here, and we leave a dedicated study of this mode to future work. The $\mu^+e^-$ mode combines three features that are well matched to the problem at hand: an electron in the initial state, a centre-of-mass energy comfortably above the mass range left open by the $B$~factories, and an initial state free of hadronic activity, which keeps the backgrounds under control. The physics potential of $\mu$TRISTAN has already attracted considerable attention, with studies covering precision observables~\cite{Hamada:2022uyn,Chen:2024tqh} and a range of beyond-the-SM scenarios~\cite{Bossi:2020yne,Das:2022mmh,Dev:2023nha,Fridell:2023gjx,Fukuda:2023yui,Lichtenstein:2023iut,Das:2024gfg,Ding:2024zaj,Calibbi:2024rcm,Das:2024kyk,Sarkar:2025bgo,Barducci:2025kuq}. To the best of our knowledge, however, the sensitivity of this machine to electrophilic ALPs and $Z'$ bosons has not been addressed so far.

We therefore study, in a model-independent effective field theory framework, a pseudoscalar and a neutral vector boson coupled exclusively to the electron, with masses between $5.5$ and $300$~GeV. In both cases, the mediator is radiated off the electron and decays back into an electron-positron pair, giving the four-lepton final state $e^-\mu^+\to e^-e^+e^-\mu^+$, whose SM counterpart proceeds through photon or $Z$ exchange. We generate signal and background events at leading order and emulate the detector response with a Gaussian momentum smearing tuned to the resolution of the ILD concept~\cite{Behnke:2013lya,ILDConceptGroup:2020sfq}. In both cases, we perform a resonance search in the invariant mass of the electron--positron pair, in which the decay pair is identified with a multivariate tagger and the reconstructed peak is fitted with a double-sided Crystal Ball function. Exclusion limits at $95\%$ CL are then derived on the signal production cross-section and on the two couplings, for four integrated luminosity configurations and for two assumptions on the systematic uncertainty, and are finally confronted with the constraints already available in the literature.

The paper is organised as follows. Section~\ref{sec:theory} introduces the effective Lagrangians for the two scenarios, together with the corresponding decay widths. Section~\ref{sec:collider} describes the collider analysis, with the ALP case treated in Section~\ref{sec:collider-alp} and the $Z'$ case in Section~\ref{sec:analysis_zpr}, and presents the projected exclusion limits. In Section~\ref{sec:interpretation}, these projections are compared with the existing bounds on electrophilic mediators. We summarise our findings in Section~\ref{sec:conclusion}.

\section{Theoretical Framework}
\label{sec:theory}

The two scenarios considered in this work extend the SM by a single neutral state that communicates with the visible sector exclusively through the electron. Rather than committing to a particular ultraviolet completion, we describe both cases by effective interactions valid below a cut-off scale $\Lambda$, assumed to lie well above the energies probed in the collider analysis of Section~\ref{sec:collider}. All couplings to the remaining SM fermions are set to zero at tree level. This choice is admittedly restrictive, but it isolates the electron coupling as the only free parameter besides the mass of the new state, and thereby allows the collider reach to be quoted without further assumptions on the flavour structure of the underlying theory. The first scenario involves a spin-zero state, generically referred to as an axion-like particle (ALP), which we take to be a pseudo-Nambu-Goldstone boson associated with the spontaneous breaking of a global $U(1)$ symmetry at the scale $\Lambda$. At the level of dimension-five operators, such a state couples to the axial current of the electron through a derivative interaction of the form $\frac{c_{ee}}{\Lambda}(\partial_\mu a)\,\bar e\gamma^\mu\gamma^5 e$, where $c_{ee}$ is the corresponding Wilson coefficient. After electroweak symmetry breaking, using the equations of motion, this operator can be written as a Yukawa-type interaction, and the Lagrangian relevant for the present analysis reads~\cite{Dine:1981rt,DiLuzio:2020wdo,Bharucha:2022lty}
\begin{equation}
\label{eq:Lalp}
\mathcal{L}_{\rm ALP}\supset\frac{1}{2}(\partial_{\mu}a)(\partial^{\mu}a)-\frac{1}{2}m_{a}^{2}a^{2}-i\,g_{aee}\,a\,\bar{e}\gamma^{5}e\,,
\end{equation}
where $m_a$ denotes the ALP mass and $g_{aee}$ the dimensionless ALP-electron coupling $g_{aee} = \frac{c_{ee}m_e}{\Lambda }$.

Since the ALP is a pseudoscalar, the interaction in Eq.~\ref{eq:Lalp} is CP-conserving. The partial decay width for $a\to e^+e^-$ is given by 
\begin{equation}
\label{eq:width_alp}
\Gamma_{a\to e^{+}e^{-}}=\frac{g_{aee}^{2}m_{a}}{8\pi}\sqrt{1-\frac{4m_{e}^{2}}{m_{a}^{2}}}\,,
\end{equation}
where the ALP decays to an electron-positron pair with a branching ratio of essentially unity~\footnote{Although Eq.~\ref{eq:Lalp} contains no tree-level coupling to photons, such an interaction is generated radiatively by an electron loop~\cite{Eberhart:2025lyu,Bauer:2017ris}, $\mathcal{L}_{a\gamma\gamma}\supset\frac{1}{4}\,g_{a\gamma\gamma}\,a\,F^{\mu\nu}\tilde{F}_{\mu\nu}$. The partial decay width is written as $\Gamma_{a\to\gamma\gamma}=\frac{m_{a}^{3}\,g_{a\gamma\gamma}^{2}}{64\pi}$. For our choice of ALP masses, the ratio $\frac{\Gamma_{\gamma \gamma}}{\Gamma_{ee}} \simeq 10^{-10} - 10^{-12}$. Hence, we can safely ignore the di-photon decay of the ALP in our analysis.}.

The second scenario replaces the pseudoscalar by a massive neutral vector boson $Z'$, associated with the breaking of an additional abelian gauge symmetry $U(1)_X$ under which the electron is charged. Keeping only the electron coupling, the corresponding Lagrangian is
\begin{equation}
\label{eq:Zpr}
\mathcal{L}_{Z'}\supset-\frac{1}{4}Z'_{\mu\nu}Z'^{\mu\nu}+\frac{1}{2}m_{Z'}^{2}Z'_{\mu}Z'^{\mu}-g_{Z'}\,\bar{e}\gamma^{\mu}e\,Z'_{\mu}\,,
\end{equation}
where $Z'_{\mu\nu}=\partial_{\mu}Z'_{\nu}-\partial_{\nu}Z'_{\mu}$ is the field strength tensor, $m_{Z'}$ is the mass of the new gauge boson and $g_{Z'}$ is its coupling to the electron current. A comment on the consistency of this construction is in order. A gauge boson coupled to the electron alone is not compatible with the cancellation of gauge anomalies, so that Eq.~\ref{eq:Zpr} should be regarded as the low-energy limit of an anomaly-free theory in which the charges of the remaining fermions are fixed by the anomaly conditions~\cite{Coloma:2022umy}. Assignments of the $L_e-L_{\mu/\tau}$ type provide a simple realisation in which the $Z'$ couples to the electron without acquiring couplings to quarks. In any such construction, the electron is embedded in an $SU(2)_L$ doublet, and the $Z'$ therefore couples to $\nu_e$ with the same strength as to the electron. We include this coupling in our analysis. It does not generate any visible signature and is thus of no direct use for the final state studied here, but it opens an invisible decay channel that reduces the rate into electrons, and it is also what renders the $Z'$ sensitive to the neutrino observables discussed in Section~\ref{sec:interpretation}. The relevant partial widths are
\begin{equation}
\label{eq:width_zpr}
\Gamma_{Z'\to e^{+}e^{-}}=\frac{g_{Z'}^{2}m_{Z'}}{12\pi}\left(1+\frac{2m_{e}^{2}}{m_{Z'}^{2}}\right)\sqrt{1-\frac{4m_{e}^{2}}{m_{Z'}^{2}}}\,,
\qquad
\Gamma_{Z'\to\nu\bar{\nu}}=\frac{g_{Z'}^{2}m_{Z'}}{24\pi}\,.
\end{equation}

Two features common to both scenarios are worth emphasising before turning to the collider analysis. First, in the parameter region of interest, namely masses between $5.5$ and $300$~GeV and couplings in the range $\mathcal{O}(10^{-5}-10^{-1})$, the new states are extremely narrow. Even for the smallest couplings probed in this work, $g\simeq6\times10^{-5}$ at a mass of a few GeV, Eqs.~\ref{eq:width_alp} and \ref{eq:width_zpr} give a total width of order $10^{-9}$~GeV, corresponding to a proper decay length $c\tau=\hbar c/\Gamma\sim10^{-7}$~m. This is orders of magnitude below the vertex resolution of any realistic detector, so that the mediator decays promptly and the $e^+e^-$ pair is reconstructed at the primary vertex. Displaced signatures, which would become relevant for substantially lighter states or weaker couplings, are therefore not considered here. Second, in both cases the new state is produced in association with the beam particles and decays back into an electron-positron pair, giving rise to the same $e^-e^+e^-\mu^+$ final state at a $\mu^+e^-$ collider. This common signature allows the two scenarios to be treated with a single analysis strategy, which we develop in the next section.

\section{Collider Analysis}
\label{sec:collider}

In this section, we investigate the collider sensitivity to a pseudoscalar and a $Z'$ with mass above $\sim 5.5~\text{GeV}$ at the proposed $\mu^+e^-$ collider, where the electron and muon beams have energies of $E_e = 30~\text{GeV}$ and $E_\mu = 1000~\text{GeV}$~\cite{Hamada:2022mua}, respectively, corresponding to a centre-of-mass energy of $\sqrt{s}=346$ GeV. As outlined in Section~\ref{sec:theory}, we first discuss the electrophilic axion-like particles (ALPs) by analyzing the $e^- \mu^+ \to e^- e^+ e^- \mu^+$ final state in Section~\ref{sec:collider-alp}. In the subsequent Section~\ref{sec:analysis_zpr}, an electrophilic $Z'$ resonance is considered in the same final state as in the ALP scenario. For event generation, we implement the ALP framework described in Section~\ref{sec:theory}, based on Eq.~\ref{eq:Lalp}, using \texttt{FeynRules}~\cite{Alloul:2013bka}, and prepare the corresponding \texttt{UFO} model file. Signal and background events are generated at the parton level, at leading order, with \texttt{MadGraph5\_aMC@NLO}~\cite{Alwall:2014hca}. Since no detector design is yet available for $\mu$TRISTAN~\cite{Hamada:2022mua,Sarkar:2025bgo}, we emulate the response by smearing the three-momentum of every final-state lepton with a Gaussian resolution function,
\begin{equation}
\label{eq:smear}
p^i_{\rm smear}=p^i\left(1+G\,\sqrt{a^{2}+\left(b\,p^i\right)^{2}+\left(c/p^i\right)^{2}}\right),
\end{equation}
where $p^i$ with $i=1,2,3$ denotes the momentum components $p_x,p_y,p_z$, and $G$ is drawn from a normal distribution of zero mean and unit variance. The parameters are fixed to $a=0$, $b=2\times10^{-5}~{\rm GeV^{-1}}$ and $c=0$, which reproduces the momentum resolution expected for the International Linear Collider (ILC)~\cite{Behnke:2013lya} in the ILD detector concept~\cite{ILDConceptGroup:2020sfq}. Since the electron mass is negligible at the energies considered here, the smeared leptons are placed back on their mass shell by setting $E=|\vec{p}\,|$, which guarantees a physical invariant mass for any reconstructed pair.

The following Asimov signal significance formula is used in this work~\cite{Cowan:2010js, Cowan:2012}:
\begin{equation}
\label{Z_sys}
{Z}_\text{A} = \sqrt{2 \left[({\rm S+B}) \log \left(\frac{({\rm S+B})({\rm B}+\sigma_{\rm B}^{2})}{{\rm B}^{2}+({\rm S+B})\sigma_{\rm B}^{2}}\right)-\frac{{\rm B}^{2}}{\sigma_{\rm B}^{2}}\log \left(1+\frac{\sigma_{\rm B}^{2} {\rm S}}{{\rm B}({\rm B}+\sigma_{\rm B}^{2})} \right) \right]} \,.
\end{equation}
The signal yield, S, and the total background yield, B, are calculated from $\sigma\cdot\epsilon\cdot\mathcal{L}$, where $\sigma$ is the cross-section of the process, $\epsilon$ is the efficiency after the collider analysis, $\mathcal{L}$ refers to the integrated luminosity, and $\sigma_{\rm B}=\sigma_{unc.}\times {\rm B}$, where $\sigma_{unc.}$ is the systematic uncertainty. In the absence of systematic uncertainty, the signal significance formula reduces to the following form:
\begin{equation}
\label{Z}
{Z}_\text{A} (\sigma_{\rm B}=0)= \sqrt{2 \left[({\rm S+B}) \log (1+\frac{\rm S}{\rm B})-{\rm S} \right]} \,.
\end{equation}

In the following subsections, we discuss the signal and background processes, the reconstruction of the electron-positron invariant mass and the extraction of the signal region, and finally present the exclusion limits on the signal production cross-section and interaction couplings for the ALP and $Z'$ scenarios.

\subsection{Electrophilic ALP scenario}
\label{sec:collider-alp}

\begin{figure}[!tb]
\begin{center}
\subfloat[]{\label{fig:11}
\includegraphics[width=.15\textwidth]{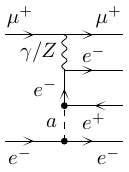}
}\qquad
\subfloat[]{\label{fig:12}
\includegraphics[height=.15\textwidth]{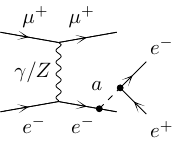}
}\qquad
\subfloat[]{\label{fig:13}
\includegraphics[width=.15\textwidth]{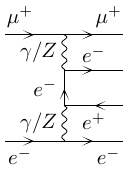}
}\qquad
\subfloat[]{\label{fig:14}
\includegraphics[height=.15\textwidth]{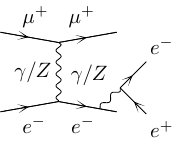}
}
\end{center}
\caption{{ The leading order (LO) Feynman diagrams for the signal process ($(a) \& (b)$) and the background process ($(c)\& (d)$) in $\mu^+e^-\to \mu^+e^+e^-e^-$ final state.}}
\label{fig:FD_eee}
\end{figure}

We begin by studying the ALP-mediated $e^- \mu^+ \to e^- e^+ e^- \mu^+$ channel. The representative Feynman diagrams for the signal process are shown in Fig.~\ref{fig:FD_eee} (a) and (b). In the ALP mass range of $m_a=(5.5-300)$ GeV, the production rate is dominated by the diagram in Fig.~\ref{fig:FD_eee}~(b). The background contribution comes from the production of the same final state, $e^- e^+ e^- \mu^+$, through SM processes mediated by a photon or a Z boson. The contributing Feynman diagrams are shown in Fig.~\ref{fig:FD_eee} (c) and (d). We impose the following generation-level requirements on the signal and background events in \texttt{MadGraph5}: the transverse momentum of the leptons must satisfy $p_{T,\ell}\geq 2$ GeV, and the leptons are restricted to the pseudo-rapidity range $|\eta_{\ell}|\leq 3.5$, where $\ell$ corresponds to an electron or a muon. Any two leptons must be separated by $\Delta R_{\ell\ell}\geq 0.2$ in the $\eta-\phi$ plane~\footnote{For two final-state particles, we define $\Delta R \equiv\sqrt{(\Delta\eta)^2 + (\Delta\phi)^2}$, with $\phi$ the azimuthal angle and the pseudorapidity related to the polar angle $\theta$ via $\eta=-\ln[\tan(\theta/2)]$.}. Since the muon beam carries a much higher energy than the electron beam, the final-state muon is produced predominantly in the forward direction. However, for the ALP mass range considered here, the final-state muon in signal events shifts toward the central region, helping reject background events. The $|\eta_{\ell}|\leq 3.5$ requirement alone reduces the background cross-section from $\sim 390$ fb (with $|\eta_{\ell}|\leq 5.0$) to $\sim 22$ fb.

We select events containing exactly three electrons and one muon in the final state. The discriminating observable of this analysis is the invariant mass of the electron--positron pair produced in the ALP decay, $m_{ee}$, which forms a narrow resonance at $m_{ee}\simeq m_a$ on top of a smooth SM continuum. We adopt $m_{ee}$ as the final discriminant because the signal remains confined to a narrow region around the resonance mass. Reconstructing $m_{ee}$ requires identifying which two of the three final-state electrons originate from the ALP. The charge assignment already resolves part of the ambiguity: the final state contains a single positron, which is necessarily a decay product of the ALP, so that the only remaining question is which of the two electrons is its partner. The other electron recoils against the ALP at the production vertex, and we call it the spectator electron. Choosing the wrong electron produces a combinatorial pair whose invariant mass is unrelated to $m_a$, and therefore populates a broad, non-resonant distribution that degrades the peak.

Simple kinematic prescriptions are not adequate over the full mass range. Pairing the positron with the angularly closest electron works well for light ALPs, whose decay products are strongly collimated, but its performance deteriorates rapidly with increasing mass: the correct pair is identified in $91\%$ of the events at $m_a=5.5$~GeV but only in $5\%$ of the events at $m_a=300$~GeV, since the opening angle of the decay products grows as the ALP becomes less boosted. Selecting the two hardest electrons shows the opposite trend, rising from $20\%$ at $m_a=5.5$~GeV to $98\%$ at $m_a=300$~GeV, and is therefore efficient only for heavy resonances. A third prescription exploits the strong longitudinal boost of the collision. Since the muon beam carries $1$~TeV against $30$~GeV for the electron beam, the whole final state is boosted towards the muon-beam direction, and the spectator electron, which recoils against the mediator at the production vertex, is emitted preferentially closer to that direction than the electron produced in the decay. Of the two electrons, the one lying further along the muon-beam direction, that is, the one with the smaller signed pseudorapidity with the $z$~axis oriented along the electron beam, is therefore identified as the spectator, and the positron is paired with the remaining electron. This ordering is considerably more stable than the two previous prescriptions, and identifies the correct pair in $63\%$ to $84\%$ of the events over the entire mass range, but it relies on a single kinematic variable and does not exploit the remaining information in the event.

We therefore identify the spectator electron with a boosted decision tree (BDT), trained with \texttt{XGBoost}~\cite{Chen:2016btl}. The problem is formulated as a binary classification over electron \emph{candidates} rather than over events: each event contributes two entries to the training sample, one for each of the two electrons, and the classifier returns for every candidate the probability $p_{\rm spec}$ that it is the spectator. Denoting by $e^-_c$ the candidate under consideration and by $e^-_o$ the other electron of the same event, the label is assigned from the generator record, $p_{\rm spec}=1$ if $e^-_c$ originates from the hard production vertex and $p_{\rm spec}=0$ if it originates from the mediator decay. At the end of the procedure, the candidate with the larger $p_{\rm spec}$ is taken as the spectator, and the positron is paired with the other electron to form the resonance candidate.

\begin{table}[tb!]
\begin{centering}
\setlength{\extrarowheight}{1mm}%
\scalebox{.95}{
\begin{tabular}{l?{1.5pt}l}
\bottomrule[1.5pt]
\multicolumn{2}{c}{\textbf{Input variables}}\tabularnewline
\bottomrule[1.5pt]
Single-lepton variables &
$E(\ell),\; p_T(\ell),\; \eta(\ell)$ \ \ for $\ell=e^-_c,\,e^-_o,\,e^+,\,\mu^+$ \tabularnewline
\cline{1-2}
Angular separations &
$\Delta R(e^-_c,e^+),\;\Delta R(e^-_c,\mu^+),\;\Delta R(e^-_c,e^-_o)$ \tabularnewline
\cline{1-2}
Pseudorapidity differences &
$\Delta\eta(e^-_c,e^+),\;\Delta\eta(e^-_c,\mu^+),\;\Delta\eta(e^-_c,e^-_o)$ \tabularnewline
\cline{1-2}
Energy difference &
$\Delta E(e^-_c,e^-_o)=E(e^-_c)-E(e^-_o)$ \tabularnewline
\cline{1-2}
Excluded by construction &
$m_{ee}$ and any variable determining it; the resonance mass \tabularnewline
\toprule[1.5pt]
\multicolumn{2}{c}{\textbf{Training configuration}}\tabularnewline
\bottomrule[1.5pt]
Algorithm & \texttt{XGBoost} gradient-boosted decision trees \tabularnewline
\cline{1-2}
Number of trees & $300$ \tabularnewline
\cline{1-2}
Maximum tree depth & $4$ \tabularnewline
\cline{1-2}
Learning rate & $0.08$ \tabularnewline
\cline{1-2}
Subsample / feature fraction & $0.9$ / $0.9$ \tabularnewline
\cline{1-2}
Loss function & binary logarithmic loss \tabularnewline
\cline{1-2}
Cross-validation & $4$-fold, grouped by event \tabularnewline
\cline{1-2}
Training sample & Two candidates per event, all mass benchmarks combined \tabularnewline
\toprule[1.5pt]
\end{tabular}
}
\par\end{centering}
\caption{Input variables and training configuration of the BDT used to identify the spectator electron. Here $e^-_c$ denotes the electron candidate being classified and $e^-_o$ the other electron of the same event. All variables are evaluated after the momentum smearing of Eq.~\ref{eq:smear}.}
\label{tab:bdt}
\end{table}

The nineteen input variables are listed in Table~\ref{tab:bdt}. They comprise the energy, the transverse momentum, and the pseudorapidity of each of the four reconstructed leptons, and a set of variables that relate the candidate electron to the rest of the event: its angular separation and pseudorapidity difference with respect to the positron, the muon, and the other electron, and its energy difference with respect to the other electron. All of them are computed after the momentum smearing of Eq.~\ref{eq:smear}, so that the classifier is trained on the same quantities that would be available experimentally. Two aspects of this choice deserve emphasis. First, the list contains only production-level kinematics: the invariant mass of the electron--positron pair, and any variable from which it could be reconstructed, is deliberately excluded, so that the tagging decision cannot sculpt the $m_{ee}$ distribution from which the signal is subsequently extracted. Second, the resonance mass is not among the inputs and the classifier is trained inclusively over all fifteen benchmarks, so that the mass-dependent transition between the kinematic regimes described above is inferred from the event itself rather than being supplied externally.

\begin{figure}[tb!]
\centering
\includegraphics[width=.55\textwidth]{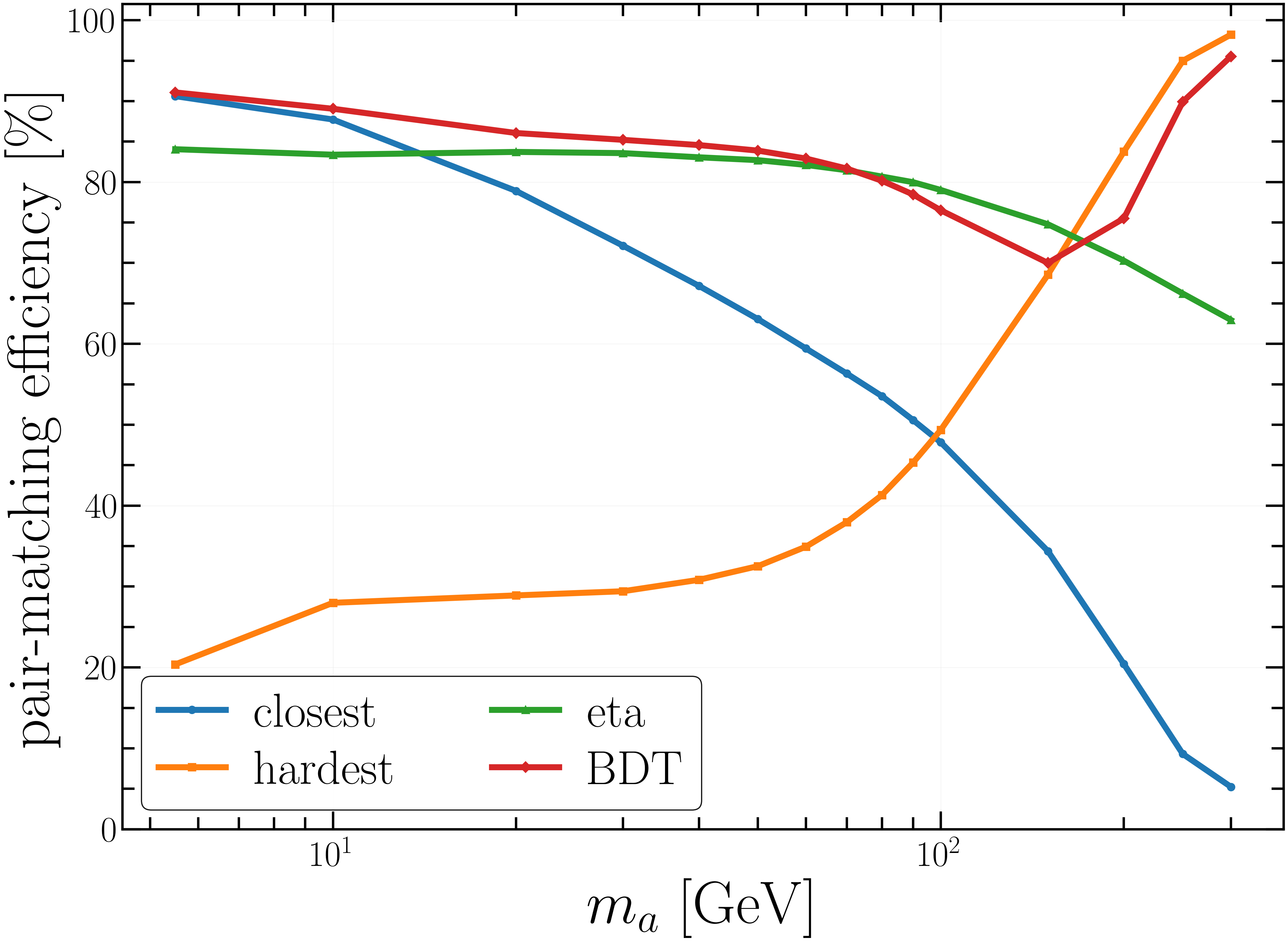}
\caption{Efficiency of correctly identifying the electron--positron pair originating from the ALP decay, as a function of the ALP mass, for the three analytic prescriptions discussed in the text and for the BDT-based tagger adopted in this analysis.}
\label{fig:pairing_eff}
\end{figure}

The hyperparameters used for the training are collected in the lower part of Table~\ref{tab:bdt}. The model is an ensemble of $300$ trees of maximum depth four, with a learning rate of $0.08$ and a logarithmic-loss objective; at each iteration, $90\%$ of the candidates and $90\%$ of the features are sampled, which limits the sensitivity to statistical fluctuations of the training sample. The performance is evaluated with a four-fold cross-validation, and the score assigned to each signal candidate is always the one obtained from the fold in which that candidate was not used for training, so that both the efficiency quoted below and the signal $m_{ee}$ distributions are free from any optimistic bias. The folds are built by grouping on the event index, ensuring that the two candidates belonging to the same event are never split between training and evaluation; without this precaution, the two entries of an event, which are strongly correlated by construction, would leak information across the folds. The background sample carries no truth label, since it contains no resonance, and is therefore tagged with a single classifier trained on the full set of signal candidates. This is the standard procedure by which a classifier trained on simulation is applied to unlabelled events: the generator record enters only in the definition of the training target, whereas the trained model acts on measured quantities alone and can be applied unchanged to data. Signal and background are thus paired by the same algorithm, so that the efficiency and the background yield quoted below are consistent with the $\mu$ and $\sigma$ that define the mass window. The resulting performance is shown in Fig.~\ref{fig:pairing_eff}. The classifier reaches an efficiency of $91\%$ at $m_a=5.5$~GeV and $96\%$ at $m_a=300$~GeV, with a minimum of $70\%$ at $m_a=150$~GeV, where the two analytic regimes cross, and neither the angular nor the momentum ordering is individually informative; averaged over the fifteen benchmarks, the correct pair is selected in $83\%$ of the events. Unlike the analytic prescriptions, whose performance varies by up to a factor of eighteen across the mass range, the classifier remains above $70\%$ everywhere, and it is the best of the four options over most of the range. It is not uniformly optimal: between $m_a\simeq80$ and $150$~GeV the pseudorapidity ordering is marginally better, and above $m_a\simeq200$~GeV the two-hardest prescription overtakes it, by up to eight percentage points at the largest masses. Since a single algorithm must be applied at all masses in order to keep the selection uniform, we retain the classifier throughout; adopting the best analytic prescription mass instead by mass would introduce a discontinuous, mass-dependent selection for a modest gain confined to the two ends of the range.

Once the pair has been identified, the invariant mass is reconstructed from the smeared momenta of the positron and of the tagged decay electron. The resulting $m_{ee}$ distribution exhibits a narrow peak centred at the ALP mass, with mildly non-Gaussian tails arising from the Breit--Wigner line shape of the resonance and from the residual fraction of incorrectly paired events. We model the peak with a double-sided Crystal Ball (DSCB) function, a Gaussian core matched continuously to two independent power-law tails, which is the standard parameterisation adopted by the LHC experiments for narrow resonances reconstructed in electromagnetic final states~\cite{ATLAS:2018tdk,ATLAS:2022abz}. Introducing the variable $t=(m_{ee}-\mu)/\sigma$, the function reads
\begin{equation}
\label{eq:dscb}
f(m_{ee})=N\times
\begin{cases}
e^{-t^{2}/2}, & -\alpha_{L}\leq t\leq\alpha_{R},\\[8pt]
e^{-\alpha_{L}^{2}/2}\left[\dfrac{\alpha_{L}}{n_{L}}\left(\dfrac{n_{L}}{\alpha_{L}}-\alpha_{L}-t\right)\right]^{-n_{L}}, & t<-\alpha_{L},\\[12pt]
e^{-\alpha_{R}^{2}/2}\left[\dfrac{\alpha_{R}}{n_{R}}\left(\dfrac{n_{R}}{\alpha_{R}}-\alpha_{R}+t\right)\right]^{-n_{R}}, & t>\alpha_{R},
\end{cases}
\end{equation}
where $\mu$ is the peak position and $\sigma$ the width of the Gaussian core, while $\alpha_{L,R}$ and $n_{L,R}$ describe the onset and the slope of the two tails. The fit is performed for each mass hypothesis separately, over a window centred on the peak, by minimising a Poissonian $\chi^{2}$ with the bin contents weighted by their statistical uncertainty.

\begin{figure}[tb!]
\centering
\includegraphics[width=.9\textwidth]{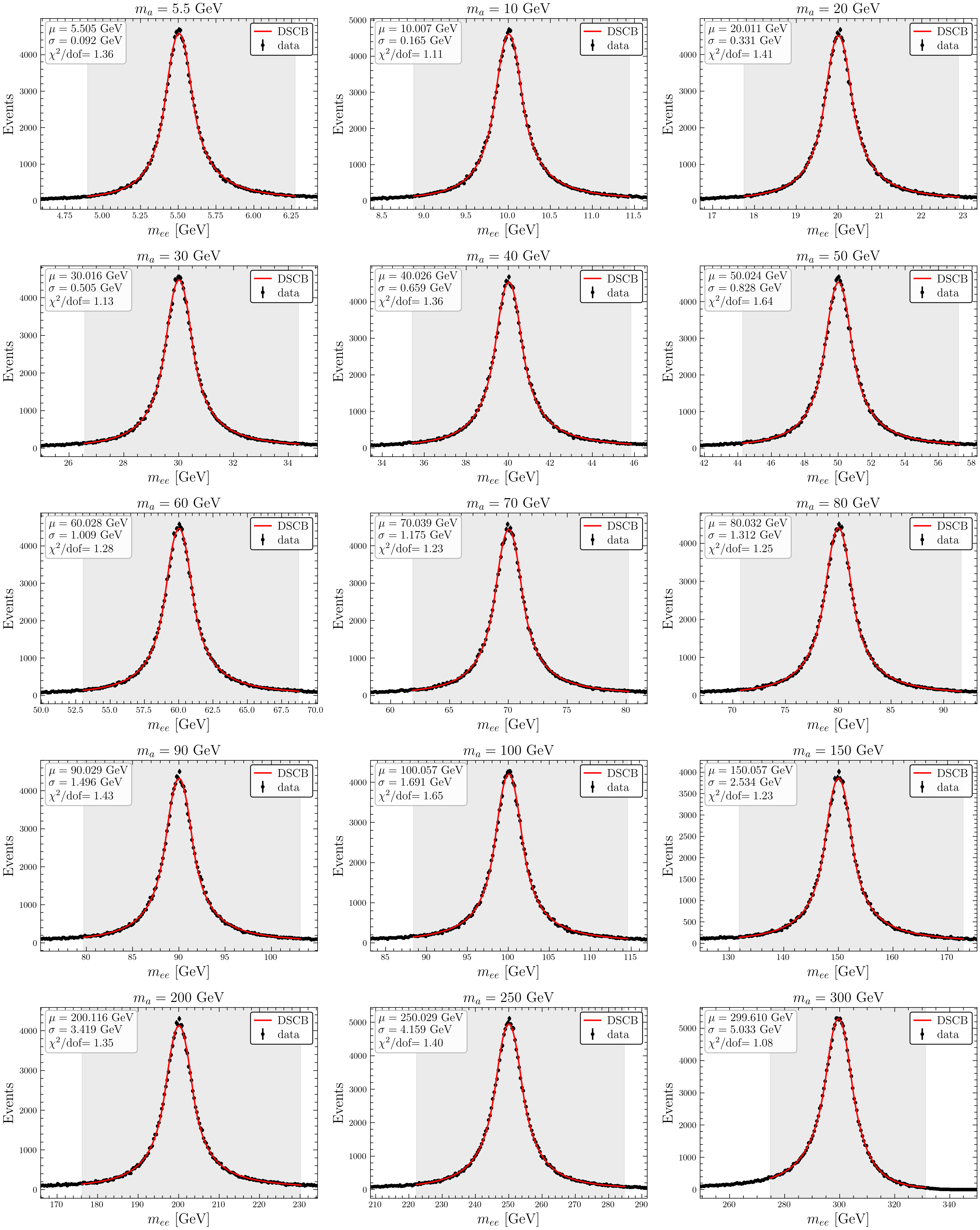}
\caption{Reconstructed $m_{ee}$ distributions for the fifteen ALP benchmark masses considered in this work. The black points show the smeared signal events after the BDT-based pairing, and the red curve is the double-sided Crystal Ball fit of Eq.~\ref{eq:dscb}. The shaded band indicates the range used in the fit, and the fitted values of $\mu$, $\sigma$ and the resulting $\chi^{2}/{\rm dof}$ are quoted in each panel.}
\label{fig:mee_fits}
\end{figure}

The fitted distributions are displayed in Fig.~\ref{fig:mee_fits} for the fifteen benchmark masses, and the corresponding parameters are collected in Table~\ref{tab:cuts_alp}. The peak position reproduces the generated mass to better than $0.15\%$ in all cases, and the relative width of the peak is stable, $\sigma/\mu=1.64-1.71\%$, over a mass range spanning a factor of more than fifty. The width of the reconstructed peak includes the natural width of the resonance at the coupling used in the simulation; since the width decreases as the square of the coupling, the signal windows derived here are conservative for the couplings to which the analysis is sensitive.

\begin{table}[tb!]
\begin{centering}
\setlength\aboverulesep{0pt}\setlength\belowrulesep{0pt}
\setlength\abovetopsep{0pt}\setlength\belowbottomsep{0pt}
\setlength{\extrarowheight}{0mm}%
\scalebox{.86}{
\begin{tabular}{c?{1.5pt}c|c|c?{1.5pt}c|c|c}
\toprule[1.pt]
\multirow{2}{*}{\makecell{$m_a$\\ $[\textrm{GeV}]$}} &
\multicolumn{3}{c?{1.5pt}}{Double-sided Crystal Ball fit} &
\multicolumn{3}{c}{\makecell{After selection \\ ($\mathcal{L}=1000~\mathrm{fb}^{-1}$)}}
\tabularnewline[0.3em]
\cline{2-7}
&
\makecell{$\mu$ \\ $[\textrm{GeV}]$} &
\makecell{$\sigma$ \\ $[\textrm{GeV}]$} &
$\chi^{2}/{\rm dof}$
&
\makecell{Signal \\ efficiency, $\epsilon$} &
\makecell{Signal yield \\ ($g_{aee}=0.1$)} &
\makecell{Background \\ yield, B}
\tabularnewline
\midrule[1.5pt]
5.5   & $5.51 \pm 0.00$    & $0.09 \pm 0.00$    &  1.4 &  0.39 &    3971 &   60 \tabularnewline[0.3em]
\cline{1-7}
10    & $10.01 \pm 0.00$   & $0.16 \pm 0.00$    &  1.1 &  0.39 &    4443 &   64 \tabularnewline[0.3em]
\cline{1-7}
20    & $20.01 \pm 0.00$   & $0.33 \pm 0.00$    &  1.4 &  0.38 &    3253 &   69 \tabularnewline[0.3em]
\cline{1-7}
30    & $30.02 \pm 0.00$   & $0.50 \pm 0.00$    &  1.1 &  0.38 &    2450 &   87 \tabularnewline[0.3em]
\cline{1-7}
40    & $40.03 \pm 0.00$   & $0.66 \pm 0.01$    &  1.4 &  0.38 &    1863 &  106 \tabularnewline[0.3em]
\cline{1-7}
50    & $50.02 \pm 0.00$   & $0.83 \pm 0.01$    &  1.6 &  0.38 &    1480 &  137 \tabularnewline[0.3em]
\cline{1-7}
60    & $60.03 \pm 0.01$   & $1.01 \pm 0.01$    &  1.3 &  0.38 &    1196 &  171 \tabularnewline[0.3em]
\cline{1-7}
70    & $70.04 \pm 0.01$   & $1.18 \pm 0.01$    &  1.2 &  0.38 &     970 &  190 \tabularnewline[0.3em]
\cline{1-7}
80    & $80.03 \pm 0.01$   & $1.31 \pm 0.01$    &  1.3 &  0.37 &     775 &  208 \tabularnewline[0.3em]
\cline{1-7}
90    & $90.03 \pm 0.01$   & $1.50 \pm 0.01$    &  1.4 &  0.37 &     643 &  489 \tabularnewline[0.3em]
\cline{1-7}
100   & $100.06 \pm 0.01$  & $1.69 \pm 0.02$    &  1.7 &  0.36 &     533 &  249 \tabularnewline[0.3em]
\cline{1-7}
150   & $150.06 \pm 0.01$  & $2.53 \pm 0.03$    &  1.2 &  0.33 &     225 &  332 \tabularnewline[0.3em]
\cline{1-7}
200   & $200.12 \pm 0.02$  & $3.42 \pm 0.03$    &  1.4 &  0.36 &     118 &  480 \tabularnewline[0.3em]
\cline{1-7}
250   & $250.03 \pm 0.02$  & $4.16 \pm 0.04$    &  1.4 &  0.42 &      68 &  381 \tabularnewline[0.3em]
\cline{1-7}
300   & $299.61 \pm 0.03$  & $5.03 \pm 0.05$    &  1.1 &  0.45 &      31 &   45 \tabularnewline[0.3em]
\bottomrule[1.pt]
\end{tabular}
}
\par\end{centering}
\caption{Parameters of the double-sided Crystal Ball fit to the reconstructed $m_{ee}$ distribution for each ALP benchmark mass: the fitted peak position $\mu$ and the width of the Gaussian core $\sigma$, quoted with their statistical uncertainties from the fit, and the reduced chi-square. All entries are rounded to two decimal places, so that an uncertainty appearing as $0.00$ indicates a value smaller than $0.005$~GeV rather than a vanishing uncertainty. The last three columns give, for the signal window $[\mu-\sigma,\,\mu+\sigma]$ and an integrated luminosity of $1000~{\rm fb}^{-1}$, the signal efficiency $\epsilon$, the signal yield ${\rm S}=\epsilon\cdot\sigma\cdot{\cal L}$ evaluated at the reference coupling $g_{aee}=0.1$, and the background yield B, the latter normalised to $\sigma_{\rm B}=22.3$~fb. Since $\sigma\propto g_{aee}^{2}$, the signal yield at any other coupling is obtained by rescaling.}
\label{tab:cuts_alp}
\end{table}

The signal region is then defined, for each benchmark, by the interval $[\mu-\sigma,\,\mu+\sigma]$ obtained from the fit. The signal efficiency $\epsilon$ is computed as the fraction of generated signal events reconstructed inside this window, and therefore includes the acceptance of the generation-level requirements, the effect of the momentum smearing, the performance of the pairing algorithm, and the fraction of the peak contained within one standard deviation. The background yield is obtained by applying the same window to the SM sample, and is normalised to $\sigma_{\rm B}=22.3$~fb and ${\cal L}=1000~{\rm fb}^{-1}$. For orientation, we also quote the corresponding signal yield,
\begin{equation}
\label{eq:syield}
{\rm S}=\epsilon\cdot\sigma(e^-\mu^+\to e^-e^+e^-\mu^+)\cdot{\cal L},
\end{equation}
evaluated at the reference coupling $g_{aee}=0.1$, with the production cross-section taken from the same calculation used below to convert the cross-section limit into a limit on the coupling. Since $\sigma\propto g_{aee}^{2}$, the yield at any other coupling follows by rescaling. All three quantities are listed in Table~\ref{tab:cuts_alp}.

The signal efficiency stays close to $0.39$ up to $m_a\simeq70$~GeV and then decreases slowly, reaching a minimum of $0.33$ at $m_a=150$~GeV before rising again to $0.45$ at $m_a=300$~GeV. This non-monotonic behaviour follows the pairing efficiency of Fig.~\ref{fig:pairing_eff}: the fraction of correctly reconstructed pairs is lowest at intermediate masses, where the angular and momentum orderings are least informative, and recovers at the upper end of the range where the decay products of a heavy, weakly boosted resonance are unambiguously the two hardest electrons. The background yield instead grows steadily with the mediator mass, simply because the width of the signal window scales with $\sigma$ and therefore admits a wider slice of the falling SM continuum. Two features deserve a comment. The yield at $m_a=90$~GeV, $489$ events, is about twice as large as at the neighbouring mass points: the signal window at this mass overlaps with the $Z\to e^+e^-$ resonance, which produces a genuine peak in the background $m_{ee}$ spectrum. At $m_a=300$~GeV, on the contrary, the background collapses to $45$ events, since the signal window approaches the kinematic limit of the machine, where the SM continuum is strongly suppressed. The signal yield, in contrast, falls monotonically above $m_a\simeq10$~GeV, from about $4.4\times10^{3}$ events to $31$ events at $m_a=300$~GeV. This decrease of more than two orders of magnitude, driven by the shrinking production cross-section rather than by the selection, is to be compared with a background that varies by less than a factor of ten: it is therefore the loss of signal, and not the growth of the background, that governs the degradation of the reach with increasing mass.

We present the final results as exclusion limits on the $e^-e^+e^-\mu^+$ signal production cross-section and on the ALP--electron coupling, $g_{aee}$, at $95\%$ confidence level (CL). To obtain the upper limit on the cross-section, we numerically solve the significance formulae of Eq.~\ref{Z_sys} and Eq.~\ref{Z}, requiring ${Z}_\text{A}> 1.96$~\footnote{This value corresponds to a two-sided $95\%$ CL for a Gaussian variable.} when the systematic uncertainty is included and ${Z}_\text{A}(\sigma_B=0)> 1.96$ in its absence. The background yield, B, corresponding to each ALP benchmark is taken from Table~\ref{tab:cuts_alp}, and the signal yield reads
$${\rm S}=\sigma(e^- \mu^+ \to e^- e^+ e^- \mu^+)\cdot\epsilon\cdot{\cal L}~,$$
where $\sigma(e^- \mu^+ \to e^- e^+ e^- \mu^+)$ is the signal production cross-section obtained from the numerical solution and $\epsilon$ is the signal efficiency of Table~\ref{tab:cuts_alp}. The highest integrated luminosity envisaged for $\mu$TRISTAN is ${\cal L}=1000~{\rm fb}^{-1}$; we consider four configurations, ${\cal L}=100$, 300, 500 and 1000 ${\rm fb}^{-1}$.

\begin{figure}[tb!]
\centering
\subfloat[]{\label{fig:3a}
\includegraphics[width=.45\textwidth]{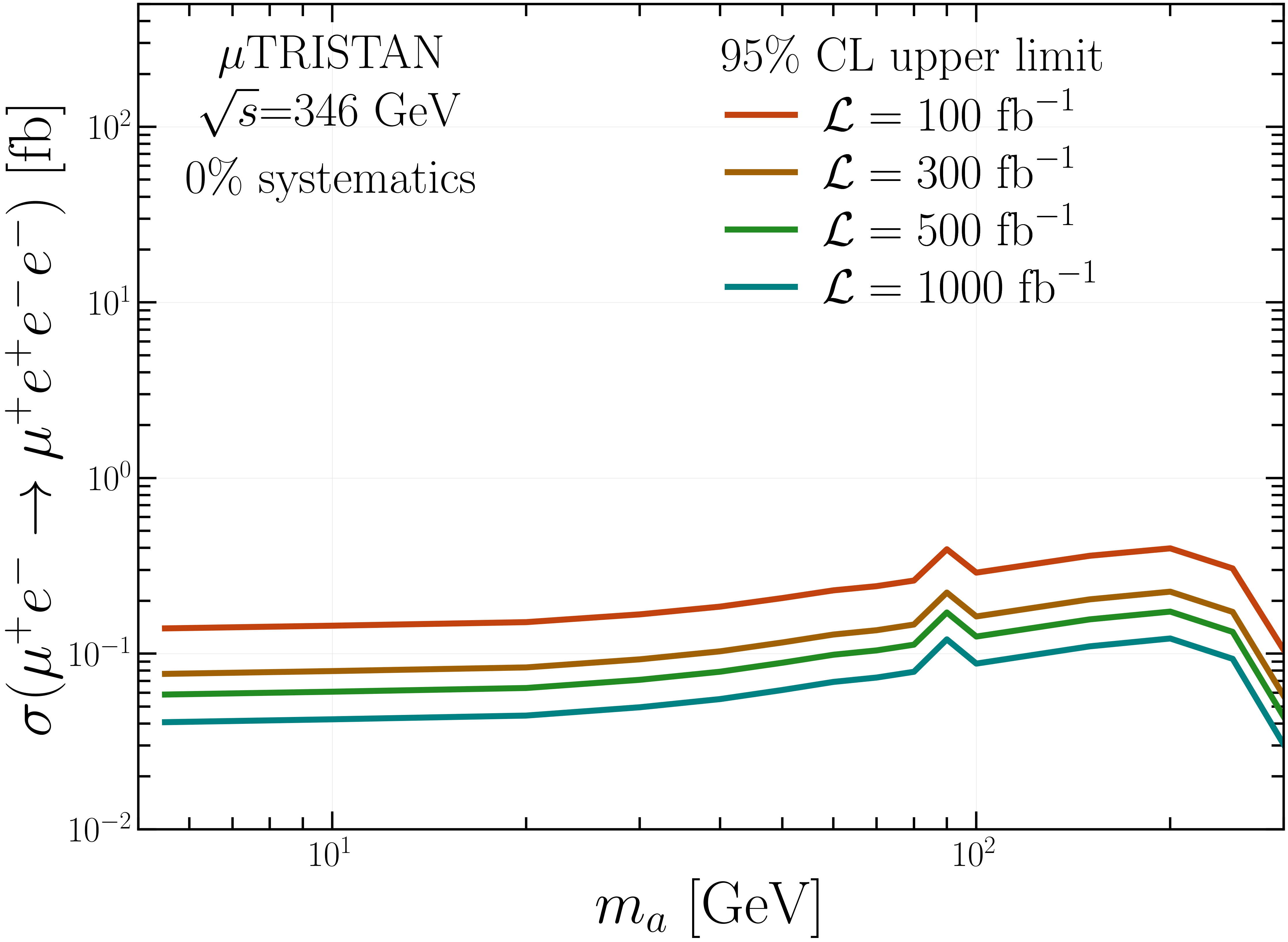}
}\qquad
\subfloat[]{\label{fig:3b}
\includegraphics[width=.45\textwidth]{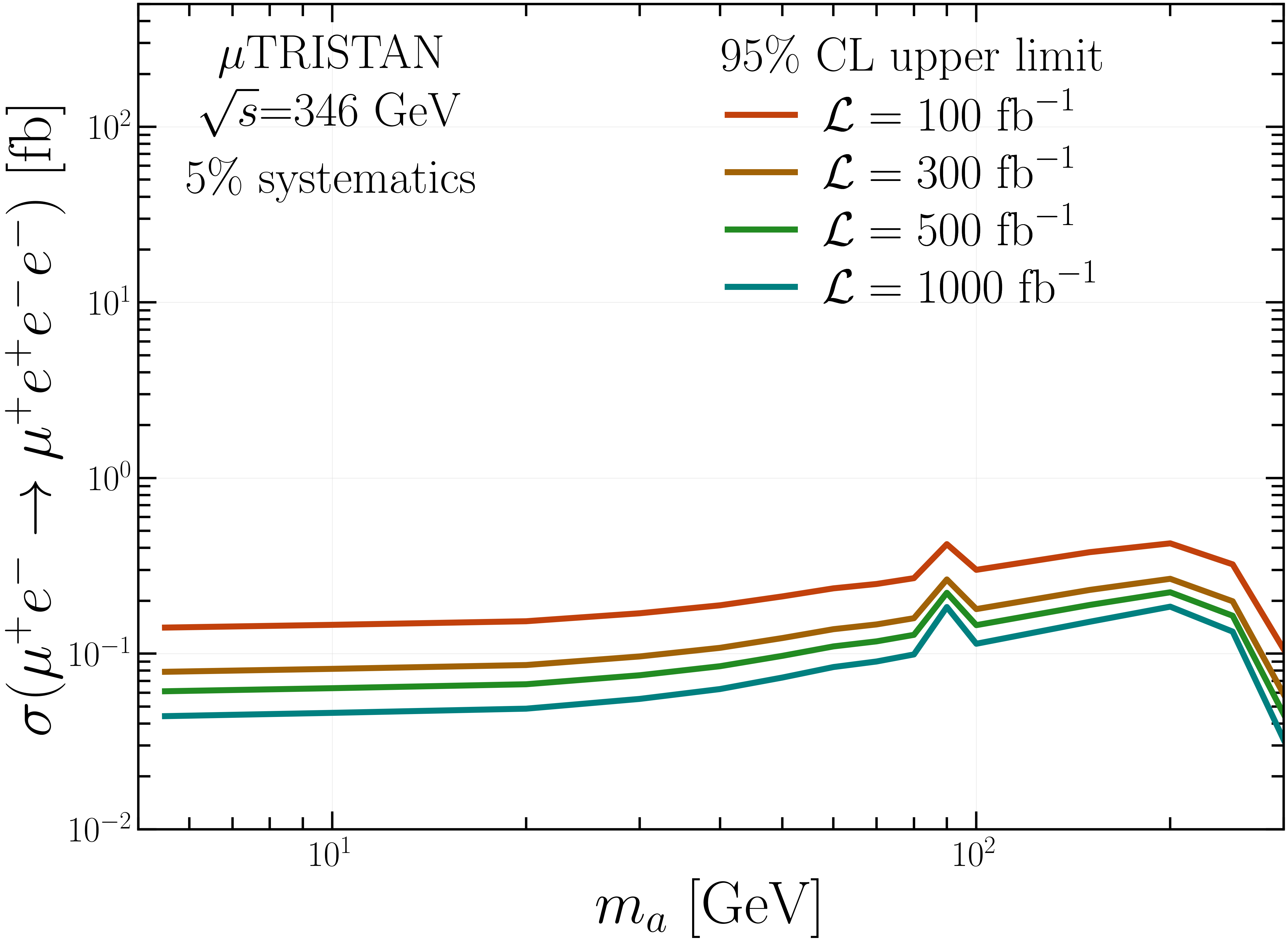}
}\\[1ex]
\subfloat[]{\label{fig:3c}
\includegraphics[width=.45\textwidth]{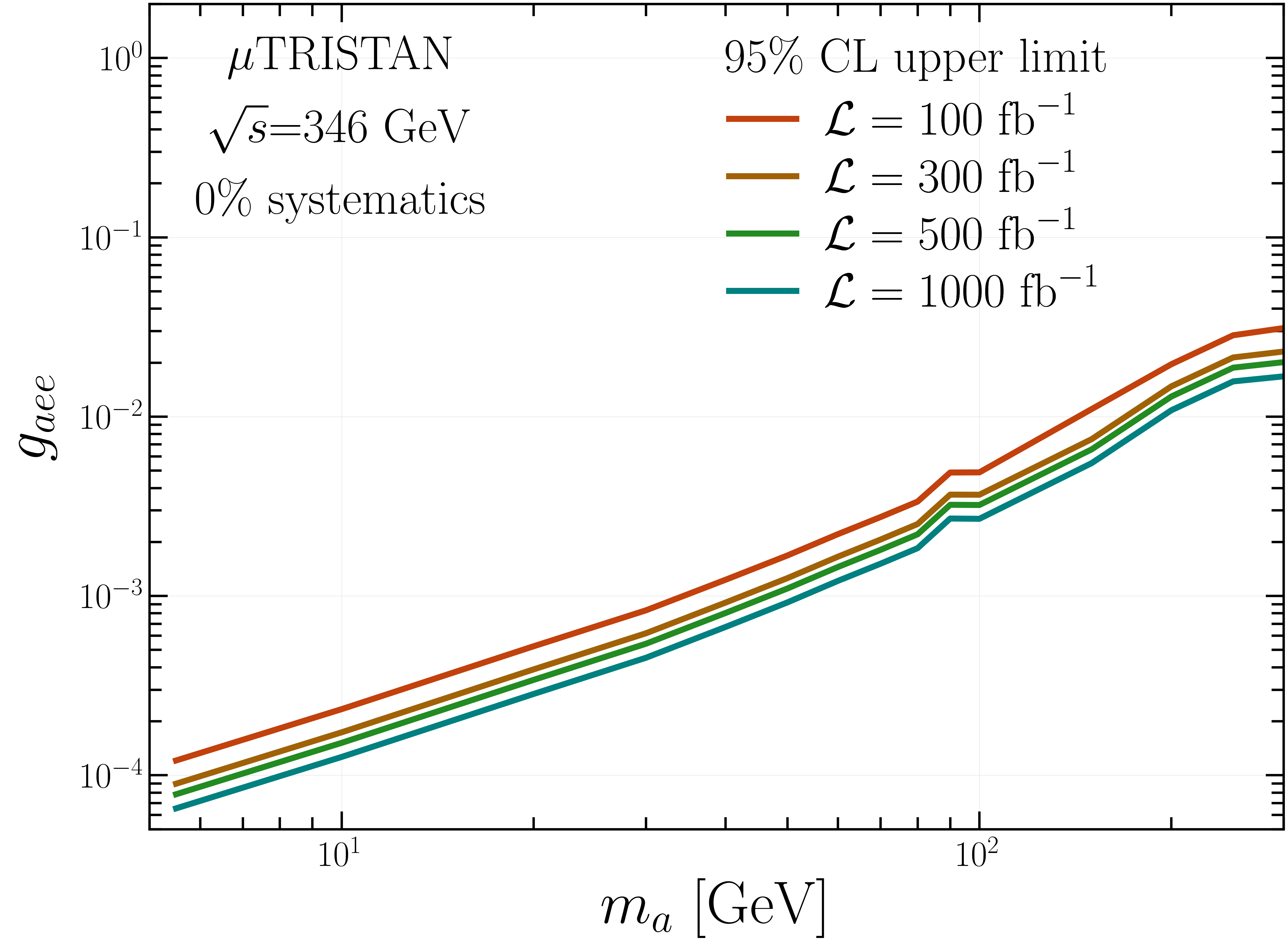}
}\qquad
\subfloat[]{\label{fig:3d}
\includegraphics[width=.45\textwidth]{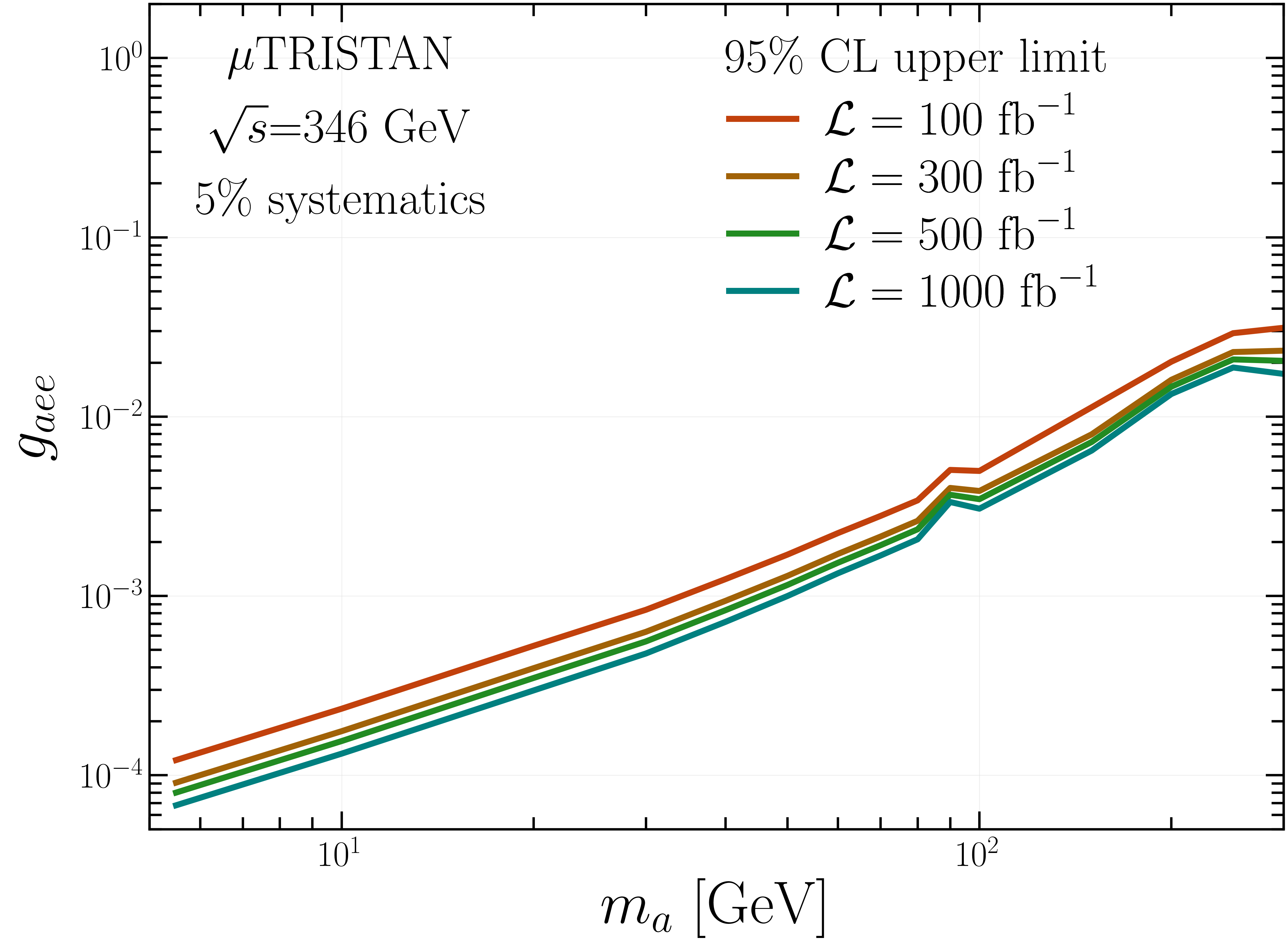}
}
\caption{Upper limit at $95\%$ CL on the signal production cross-section, $\sigma(\mu^+e^-\to \mu^+e^+e^-e^-)$ as a function of ALP mass, $m_a$ (a) with null systematics and (b) with $5\%$ systematics, and the ALP-electron coupling, $g_{aee}$ (c) with null systematics and (d) with $5\%$ systematics. The different colors correspond to variation with integrated luminosity.}
\label{fig:ul_gaee_alp}
\end{figure}

The $95\%$ CL exclusion limits on $\sigma(e^- \mu^+ \to e^- e^+ e^- \mu^+)$ are displayed in Fig.~\ref{fig:ul_gaee_alp}~(a) for null systematic uncertainty and in Fig.~\ref{fig:ul_gaee_alp}~(b) for a systematic uncertainty of $5\%$. The limits are strongest for ${\cal L}=1000~{\rm fb}^{-1}$, where the excluded cross-section ranges from $0.03$~fb at $m_a=5.5$~GeV to $0.12$~fb at $m_a=200$~GeV without systematics, and from $0.03$~fb to $0.19$~fb once a $5\%$ uncertainty is included. The cross-section limit is remarkably flat below $m_a\simeq50$~GeV, reflecting the combination of an almost constant signal efficiency with a background yield that grows only slowly in this regime, and it degrades gradually towards higher masses as the signal window widens. Two departures from this smooth behaviour are visible and can be traced directly to Table~\ref{tab:cuts_alp}: a local worsening at $m_a=90$~GeV, caused by the $Z\to e^+e^-$ contribution entering the signal window, and a pronounced improvement at $m_a=300$~GeV, where the background almost vanishes, and the limit returns to $0.03$~fb. The impact of the systematic uncertainty is modest throughout, and is largest where the background yield is largest.

The corresponding limits on the ALP--electron coupling are shown in Fig.~\ref{fig:ul_gaee_alp}~(c) and (d); couplings above the curves are excluded at $95\%$ CL for the given luminosity and systematic uncertainty. For ${\cal L}=1000~{\rm fb}^{-1}$ and without systematics, the reach varies between $g_{aee}\lesssim6.5\times10^{-5}$ and $2.7\times10^{-3}$ for $5.5~{\rm GeV}\leq m_a\leq100$~GeV, and between $5.5\times10^{-3}$ and $1.7\times10^{-2}$ for $150~{\rm GeV}\leq m_a\leq300$~GeV. Including a $5\%$ systematic uncertainty relaxes these values to $6.8\times10^{-5}-3.3\times10^{-3}$ and $6.5\times10^{-3}-1.9\times10^{-2}$, respectively; the degradation never exceeds $25\%$ and is largest where the background yield is largest.

The projected limit on the coupling degrades monotonically with the ALP mass, in contrast with the much flatter behaviour of the cross-section limit. The two are related by $\sigma\propto g_{aee}^{2}$, so that the coupling reach is governed by the rapid fall of the ALP production cross-section rather than by the performance of the analysis: at $\sqrt{s}=346$~GeV, producing a resonance of a few hundred GeV together with the recoiling electron and the forward muon is increasingly suppressed by the available phase space. The signal efficiency, which varies only between $0.33$ and $0.45$ across the entire mass range, plays a comparatively minor role. This is illustrated by the behaviour at $m_a=300$~GeV: although the cross-section limit improves there by a factor of about four with respect to $m_a=200$~GeV, owing to the strong suppression of the background, the corresponding limit on $g_{aee}$ still weakens, because the loss in production rate between the two masses is larger than the gain in sensitivity.

\subsection{Electrophilic $Z'$ scenario}
\label{sec:analysis_zpr}

\begin{figure}[!tb]
\begin{center}
\subfloat[]{\label{fig:4a}
\includegraphics[width=.15\textwidth]{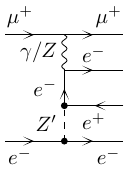}
}\qquad
\subfloat[]{\label{fig:4b}
\includegraphics[height=.15\textwidth]{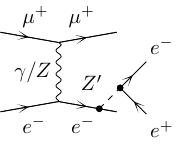}
}
\end{center}
\caption{The leading-order (LO) Feynman diagrams for the signal process in the case of the $Z'$ scenario. The Feynman diagrams for the background process are the same as in Fig.~\ref{fig:FD_eee}.}
\label{fig:FD_zpr}
\end{figure}

We now turn to the electrophilic $Z'$ scenario in the same $e^-\mu^+\to e^-e^+e^-\mu^+$ final state discussed in Section~\ref{sec:collider-alp}. The representative Feynman diagrams for the $Z'$ signal process are shown in Fig.~\ref{fig:FD_zpr}~(a) and (b), which share the same topology as the corresponding ALP diagrams in Fig.~\ref{fig:FD_eee}~(a) and (b), with the $Z'$ vector boson replacing the ALP at the relevant vertices. As in the ALP case, for the $Z'$ mass range considered in this work, $m_{Z'}=(5.5-300)$ GeV, the production is dominated by the diagram in Fig.~\ref{fig:FD_zpr}~(b). The SM background is identical to that used in the ALP analysis, arising from the production of the same $e^-e^+e^-\mu^+$ final state mediated by a photon or a $Z$ boson, with the contributing diagrams shown in Fig.~\ref{fig:FD_eee}~(c) and (d). We therefore impose the same generation-level requirements on the signal and background events: $p_{T,\ell}\geq 2$ GeV, $|\eta_\ell|\leq 3.5$, and $\Delta R_{\ell\ell}\geq 0.2$ for any two leptons in the event, where $\ell$ again corresponds to an electron or a muon. Since the muon beam carries a much higher energy than the electron beam, the final-state muon in the $Z'$ signal similarly moves toward the central pseudorapidity region for the $Z'$ masses considered here, as was observed for the ALP signal.

\begin{figure}[tb!]
\centering
\includegraphics[width=.55\textwidth]{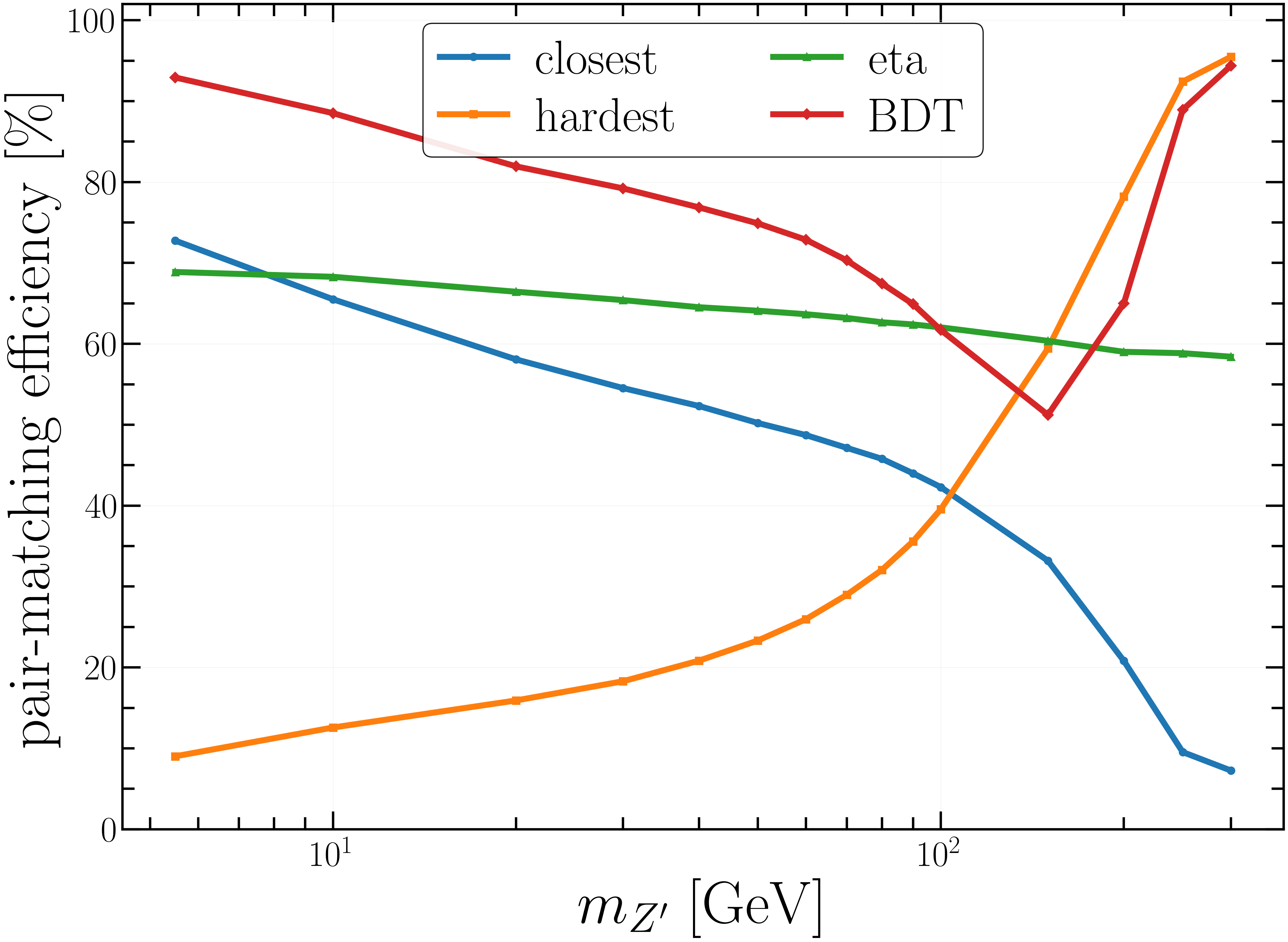}
\caption{Efficiency of correctly identifying the electron--positron pair originating from the $Z'$ decay, as a function of the $Z'$ mass, for the three analytic prescriptions discussed in Section~\ref{sec:collider-alp} and for the BDT-based tagger adopted in this analysis.}
\label{fig:pairing_eff_zpr}
\end{figure}

We select events containing exactly three electrons and one muon in the final state and follow the same strategy as in the ALP case, using the invariant mass of the electron--positron pair, $m_{ee}$, as the final discriminant.~\footnote{Because the ($Z'$) signal and the SM $Z$-mediated background populate the same final state, the corresponding amplitudes interfere. We have evaluated the interference contribution explicitly and find that, after imposing the narrow invariant-mass selection around the ($Z'$) resonance, its effect on the derived coupling limits remains below 0.4\%~\cite{Dicus:1987fk,Dixon:2003yb,Martin:2012xc,Carena:2016npr,Kahlhoefer:2019vhz}. The signal--background interference is therefore numerically negligible for the present analysis and is omitted henceforth.} The charge structure of the final state is identical to the one discussed in Section~\ref{sec:collider-alp}: a single positron, which is necessarily a decay product of the resonance, and two electrons, only one of which originates from the $Z'$ decay. The spectator electron is therefore identified with the same BDT-based tagger, trained on production kinematics alone and with the invariant mass of the pair excluded from the list of input features. The performance of the tagger follows the same pattern as in the ALP case, as expected from the fact that the kinematics of the $e^-e^+e^-\mu^+$ final state are governed primarily by the mass of the intermediate resonance rather than by its spin: the correct pair is selected in $93\%$ of the events at $m_{Z'}=5.5$~GeV and in $94\%$ at $m_{Z'}=300$~GeV, with a minimum of $51\%$ at $m_{Z'}=150$~GeV, where the angular and momentum orderings cross over. The corresponding efficiencies are shown in Fig.~\ref{fig:pairing_eff_zpr}.

\begin{figure}[tb!]
\centering
\includegraphics[width=.9\textwidth]{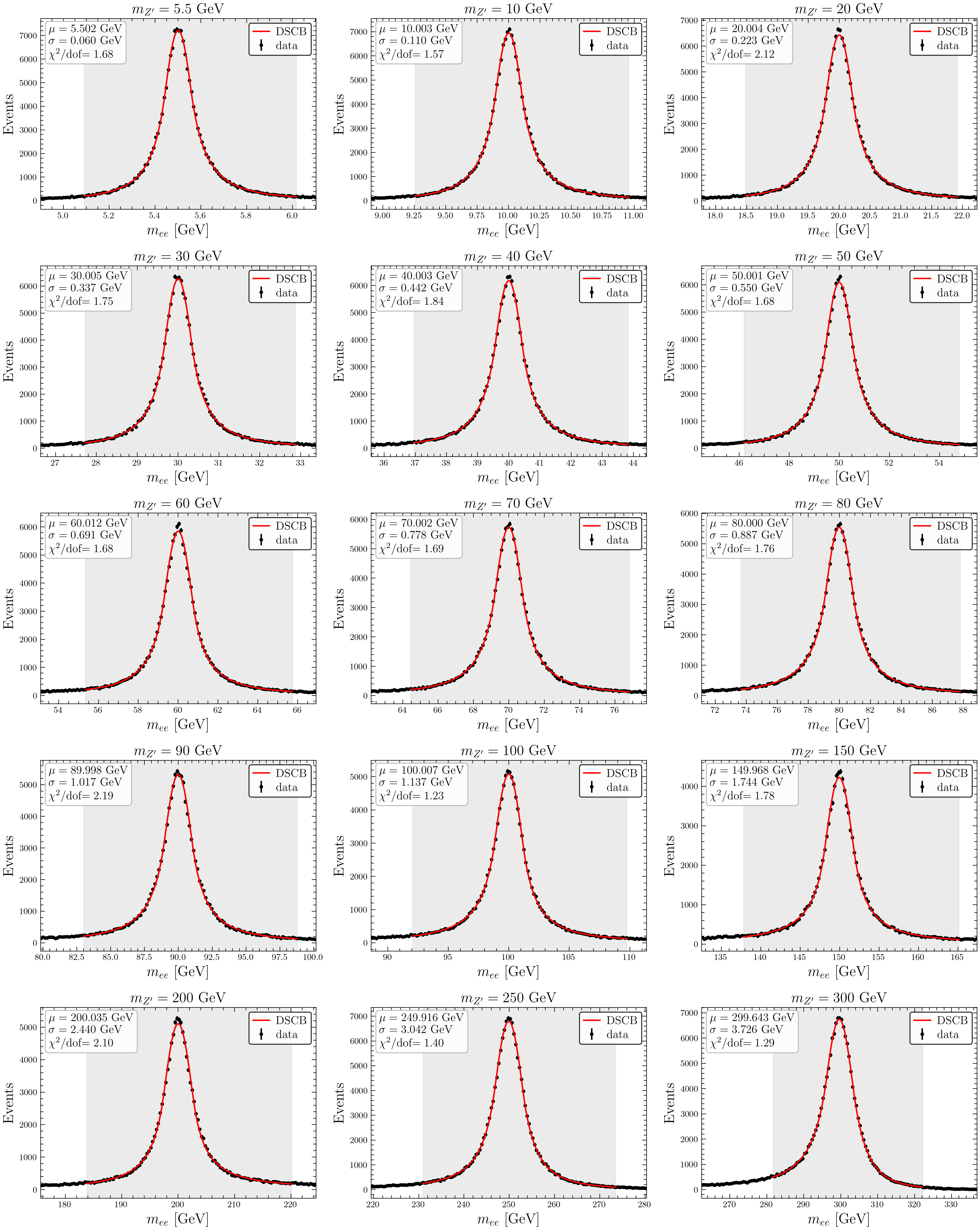}
\caption{Reconstructed $m_{ee}$ distributions for the fifteen $Z'$ benchmark masses considered in this work. The black points show the smeared signal events after the BDT-based pairing, and the red curve is the double-sided Crystal Ball fit of Eq.~\ref{eq:dscb}. The shaded band indicates the range used in the fit, and the fitted values of $\mu$, $\sigma$ and the resulting $\chi^{2}/{\rm dof}$ are quoted in each panel.}
\label{fig:mee_fits_zpr}
\end{figure}

The reconstructed $m_{ee}$ distributions are shown in Fig.~\ref{fig:mee_fits_zpr}, together with the double-sided Crystal Ball fits of Eq.~\ref{eq:dscb}, and the fitted parameters are collected in Table~\ref{tab:cuts_zprime}. As in the ALP case, the fitted peak position reproduces the generated mass to better than $0.15\%$ over the entire range, and the relative width of the peak is stable, $\sigma/\mu=1.09-1.24\%$, with only a mild increase above $m_{Z'}=150$~GeV.

\begin{table}[tb!]
\begin{centering}
\setlength{\extrarowheight}{0mm}%
\setlength\aboverulesep{0pt}\setlength\belowrulesep{0pt}
\setlength\abovetopsep{0pt}\setlength\belowbottomsep{0pt}
\scalebox{.86}{
\begin{tabular}{c?{1.5pt}c|c|c?{1.5pt}c|c|c}
\bottomrule
\multirow{2}{*}{\makecell{$m_{Z'}$\\ $[\textrm{GeV}]$}} &
\multicolumn{3}{c?{1.5pt}}{Double-sided Crystal Ball fit} &
\multicolumn{3}{c}{\makecell{After selection \\ ($\mathcal{L}=1000~\mathrm{fb}^{-1}$)}}
\tabularnewline[0.3em]
\cline{2-7}

&
\makecell{$\mu$ \\ $[\textrm{GeV}]$} &
\makecell{$\sigma$ \\ $[\textrm{GeV}]$} &
$\chi^{2}/{\rm dof}$
&
\makecell{Signal \\ efficiency, $\epsilon$} &
\makecell{Signal yield \\ ($g_{Z'}=0.005$)} &
\makecell{Background \\ yield, B}
\tabularnewline
\midrule[1.5pt]
5.5   & $5.50 \pm 0.00$    & $0.06 \pm 0.00$    &  1.7 &  0.40 &     118 &   58 \tabularnewline[0.3em]
\cline{1-7}
10    & $10.00 \pm 0.00$   & $0.11 \pm 0.00$    &  1.6 &  0.39 &      99 &   56 \tabularnewline[0.3em]
\cline{1-7}
20    & $20.00 \pm 0.00$   & $0.22 \pm 0.00$    &  2.1 &  0.37 &      56 &   56 \tabularnewline[0.3em]
\cline{1-7}
30    & $30.01 \pm 0.00$   & $0.34 \pm 0.00$    &  1.7 &  0.36 &      35 &   67 \tabularnewline[0.3em]
\cline{1-7}
40    & $40.00 \pm 0.00$   & $0.44 \pm 0.00$    &  1.8 &  0.35 &      24 &   81 \tabularnewline[0.3em]
\cline{1-7}
50    & $50.00 \pm 0.00$   & $0.55 \pm 0.01$    &  1.7 &  0.34 &      17 &   94 \tabularnewline[0.3em]
\cline{1-7}
60    & $60.01 \pm 0.00$   & $0.69 \pm 0.01$    &  1.7 &  0.34 &      13 &  117 \tabularnewline[0.3em]
\cline{1-7}
70    & $70.00 \pm 0.00$   & $0.78 \pm 0.01$    &  1.7 &  0.33 &     9.8 &  120 \tabularnewline[0.3em]
\cline{1-7}
80    & $80.00 \pm 0.01$   & $0.89 \pm 0.01$    &  1.8 &  0.31 &     7.3 &  129 \tabularnewline[0.3em]
\cline{1-7}
90    & $90.00 \pm 0.01$   & $1.02 \pm 0.01$    &  2.2 &  0.31 &     5.7 &  375 \tabularnewline[0.3em]
\cline{1-7}
100   & $100.01 \pm 0.01$  & $1.14 \pm 0.01$    &  1.2 &  0.30 &     4.8 &  144 \tabularnewline[0.3em]
\cline{1-7}
150   & $149.97 \pm 0.01$  & $1.74 \pm 0.02$    &  1.8 &  0.25 &     1.5 &  162 \tabularnewline[0.3em]
\cline{1-7}
200   & $200.03 \pm 0.01$  & $2.44 \pm 0.02$    &  2.1 &  0.32 &    0.89 &  275 \tabularnewline[0.3em]
\cline{1-7}
250   & $249.92 \pm 0.02$  & $3.04 \pm 0.03$    &  1.4 &  0.42 &    0.49 &  253 \tabularnewline[0.3em]
\cline{1-7}
300   & $299.64 \pm 0.02$  & $3.73 \pm 0.03$    &  1.3 &  0.43 &    0.19 &   32 \tabularnewline[0.3em]
\bottomrule[1.pt]
\end{tabular}
}
\par\end{centering}
\caption{Parameters of the double-sided Crystal Ball fit to the reconstructed $m_{ee}$ distribution for each $Z'$ benchmark mass: the fitted peak position $\mu$ and the width of the Gaussian core $\sigma$, quoted with their statistical uncertainties from the fit, and the reduced chi-square. All entries are rounded to two decimal places, so that an uncertainty appearing as $0.00$ indicates a value smaller than $0.005$~GeV rather than a vanishing uncertainty. The last three columns give, for the signal window $[\mu-\sigma,\,\mu+\sigma]$ and an integrated luminosity of $1000~{\rm fb}^{-1}$, the signal efficiency $\epsilon$, the signal yield ${\rm S}=\epsilon\cdot\sigma\cdot{\cal L}$ evaluated at the reference coupling $g_{Z'}=0.005$, and the background yield B, the latter normalised to $\sigma_{\rm B}=22.3$~fb. Since $\sigma\propto g_{Z'}^{2}$, the signal yield at any other coupling is obtained by rescaling. Note that the reference coupling differs from the one adopted for the ALP in Table~\ref{tab:cuts_alp}.}
\label{tab:cuts_zprime}
\end{table}

The signal region is again defined by the interval $[\mu-\sigma,\,\mu+\sigma]$, and the corresponding signal efficiency, signal yield and background yield are listed in the last three columns of Table~\ref{tab:cuts_zprime}. The signal yield is obtained from Eq.~\ref{eq:syield} at the reference coupling $g_{Z'}=0.005$, chosen so that the yields span a convenient range, and again decreases monotonically with the mediator mass, by almost three orders of magnitude between $m_{Z'}=5.5$ and $300$~GeV. The efficiency decreases smoothly from $0.40$ at $m_{Z'}=5.5$~GeV to a minimum of $0.25$ at $m_{Z'}=150$~GeV, and then recovers to $0.43$ at $m_{Z'}=300$~GeV, mirroring the shape of the pairing efficiency. The background yield behaves in the same way as before, growing from $58$ events at $m_{Z'}=5.5$~GeV to $275$ events at $m_{Z'}=200$~GeV as the window widens, with the same two exceptions: an enhancement to $375$ events at $m_{Z'}=90$~GeV, where the window overlaps the $Z\to e^+e^-$ peak of the SM background, and a collapse to $32$ events at $m_{Z'}=300$~GeV, close to the kinematic limit of the machine. Compared with the ALP case, the background admitted by the window is smaller by about a factor of $1.5$ over most of the mass range, which follows directly from the narrower resonance.

We follow the same procedure outlined in Section~\ref{sec:collider-alp} to obtain the $95\%$ CL exclusion limits on the $e^-e^+e^-\mu^+$ signal production cross-section and on the $Z'$-electron coupling, $g_{Z'}$: we numerically solve the significance formulae of Eq.~\ref{Z_sys} and Eq.~\ref{Z}, requiring ${Z}_\text{A}> 1.96$ when the systematic uncertainty is included and ${Z}_\text{A}(\sigma_B=0)> 1.96$ in its absence. The background yield, B, corresponding to each $Z'$ benchmark is taken from Table~\ref{tab:cuts_zprime}, and the signal yield is given by ${\rm S}=\sigma(e^- \mu^+ \to e^- e^+ e^- \mu^+)\cdot\epsilon\cdot{\cal L}$, where $\epsilon$ is the signal efficiency of the same table. As in the ALP scenario, we consider the four integrated luminosity configurations ${\cal L}=100$, 300, 500 and 1000 ${\rm fb}^{-1}$.

\begin{figure}[tb!]
\centering
\subfloat[]{\label{fig:5a}
\includegraphics[width=.45\textwidth]{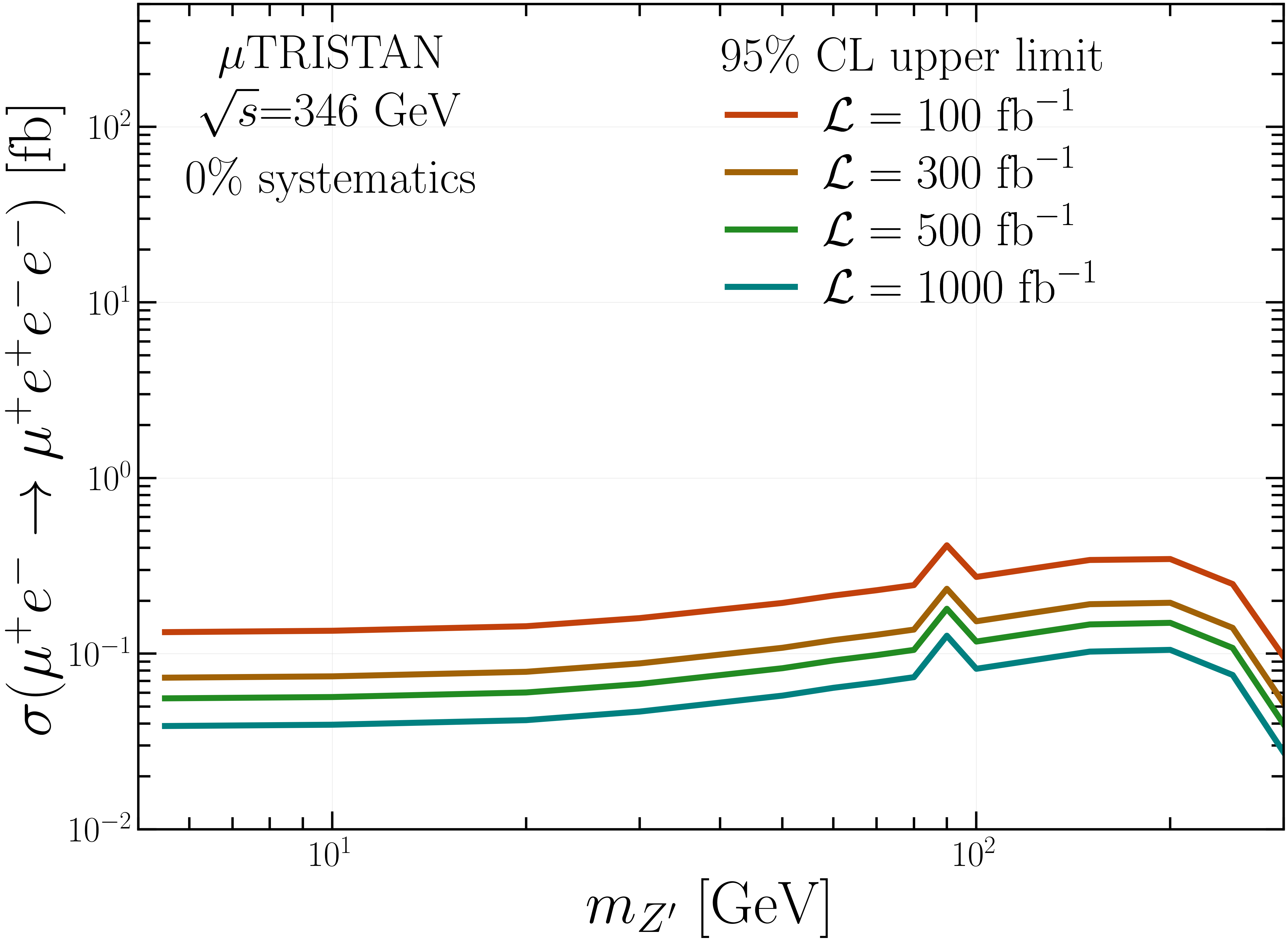}
}\qquad
\subfloat[]{\label{fig:5b}
\includegraphics[width=.45\textwidth]{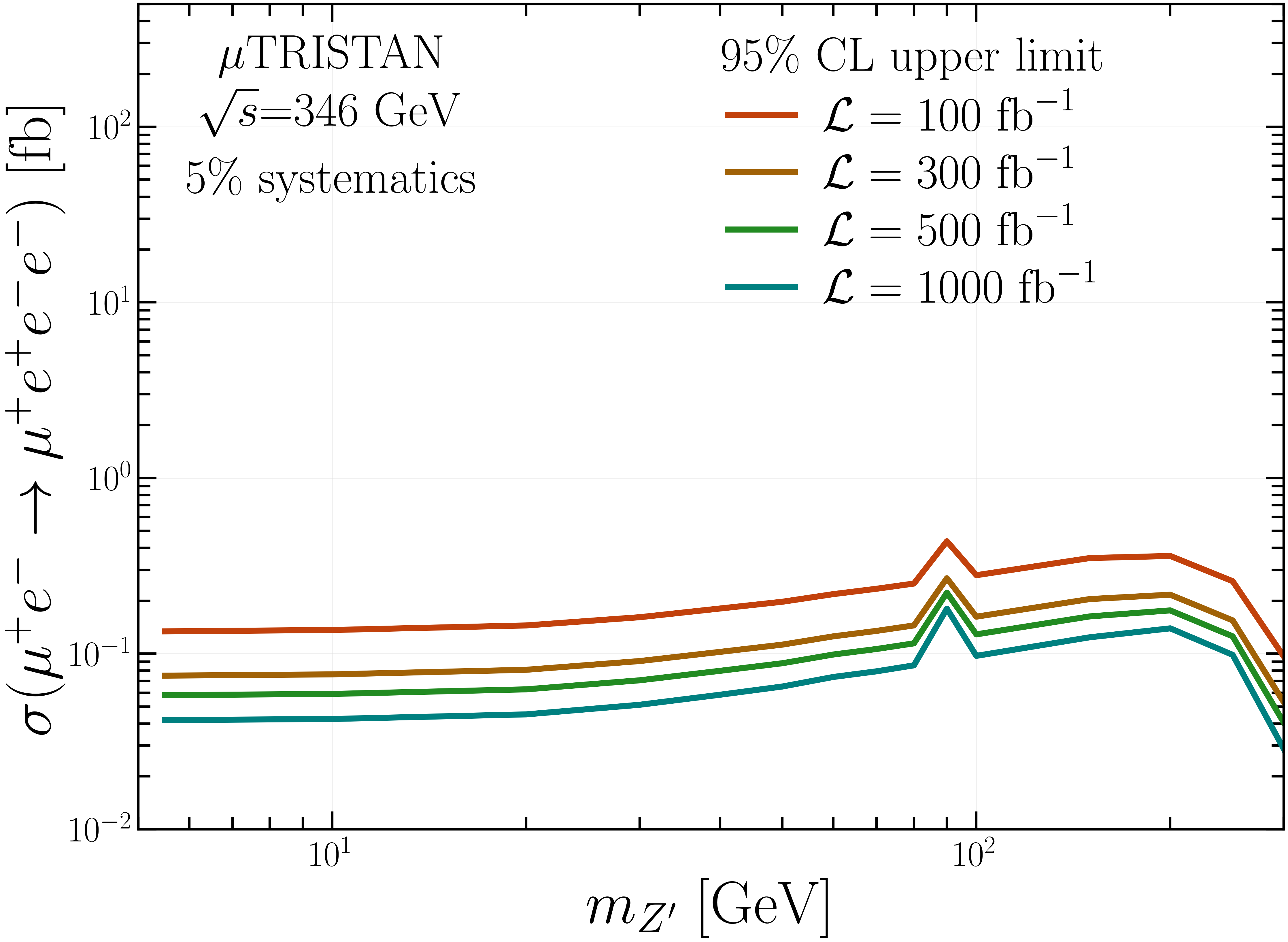}
}\\[1ex]
\subfloat[]{\label{fig:5c}
\includegraphics[width=.45\textwidth]{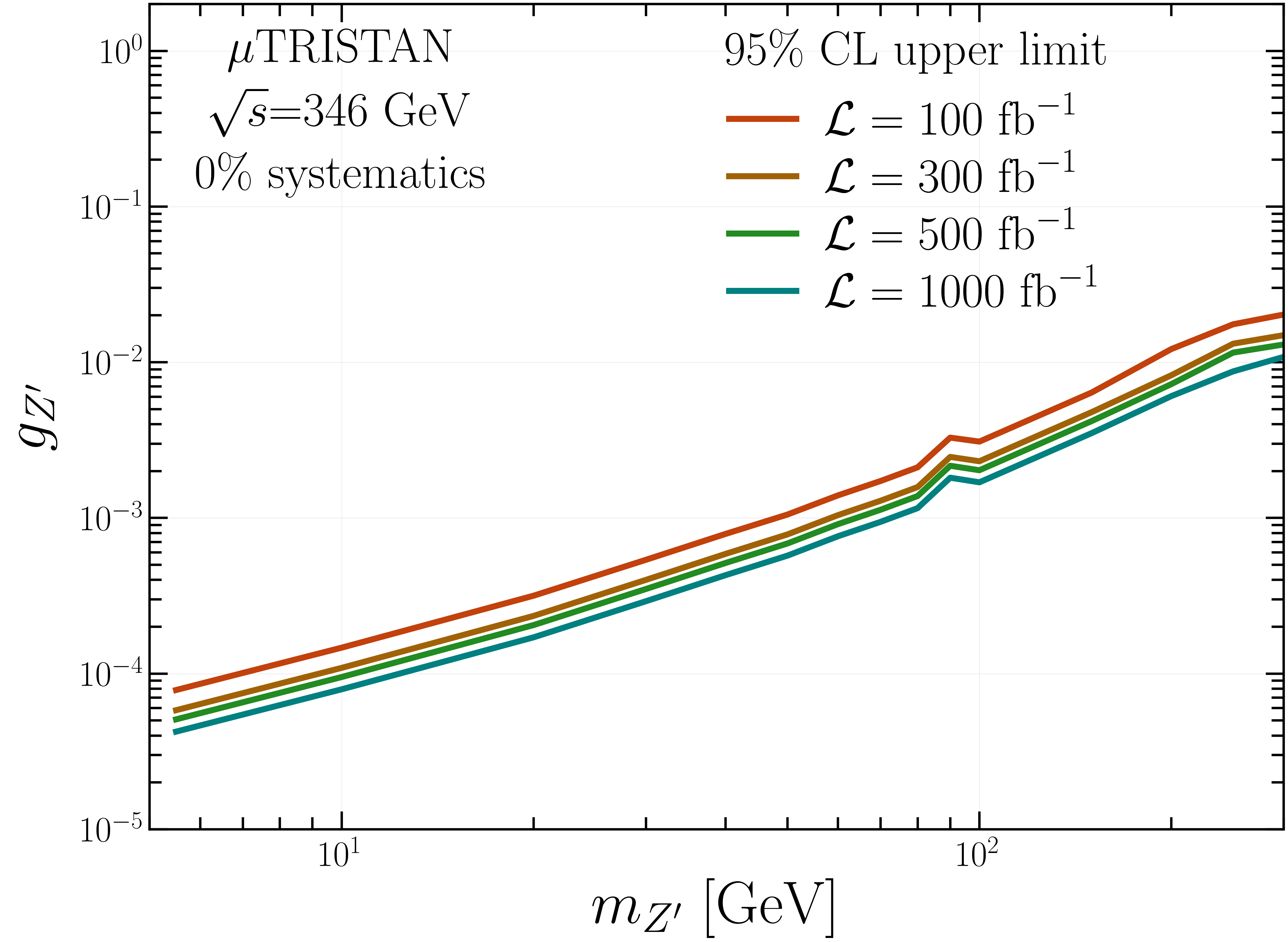}
}\qquad
\subfloat[]{\label{fig:5d}
\includegraphics[width=.45\textwidth]{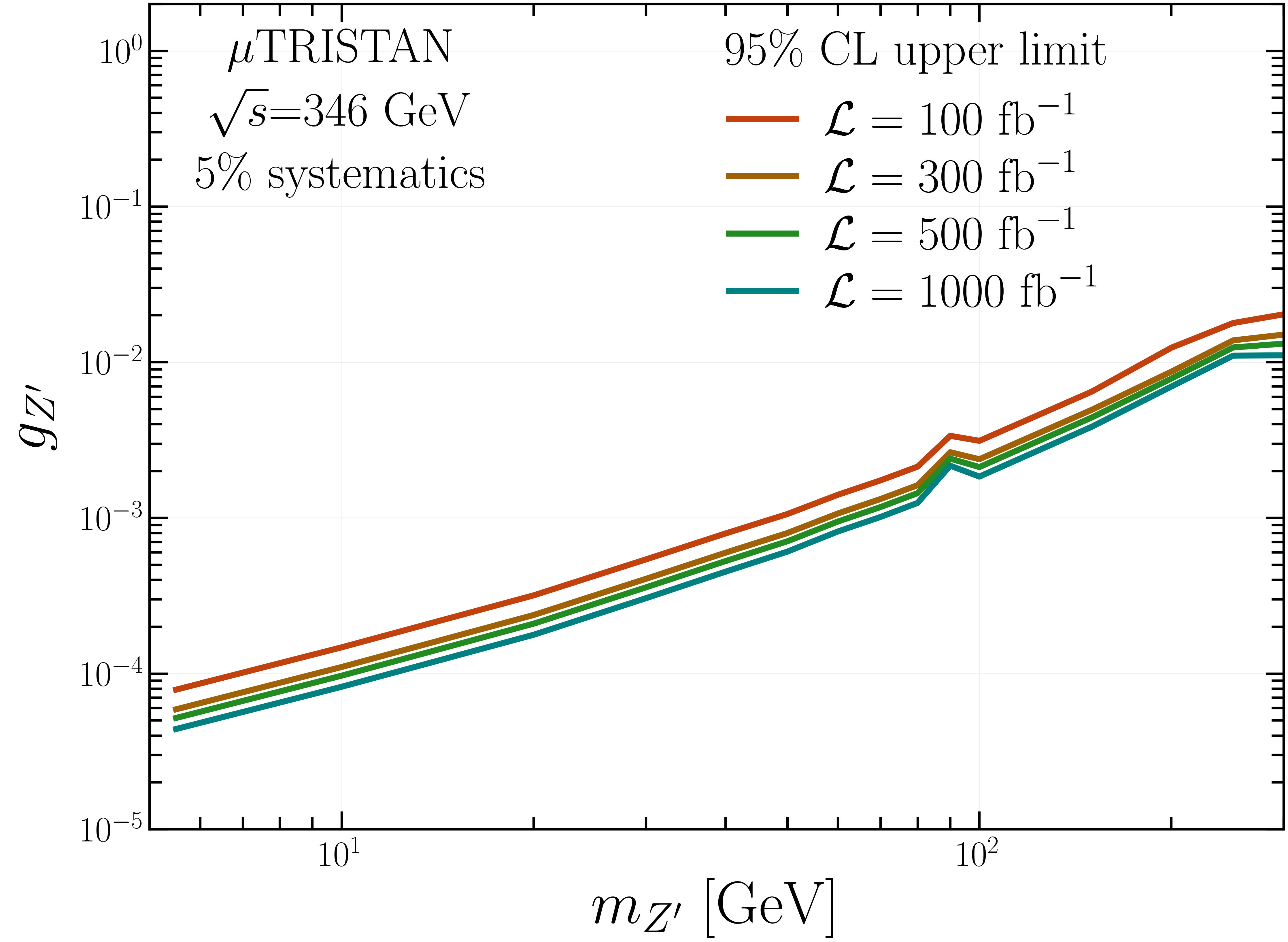}
}
\caption{Upper limit at $95\%$ CL on the signal production cross-section, $\sigma(\mu^+e^-\to \mu^+e^+e^-e^-)$ as a function of $Z'$ mass, $m_{\rm Z'}$ (a) with null systematics and (b) with $5\%$ systematics, and the $Z'$-electron coupling, $g_{\rm Z'}$ (c) with null systematics and (d) with $5\%$ systematics. The different colors correspond to variation with integrated luminosity.}
\label{fig:ul_gz}
\end{figure}

The $95\%$ CL exclusion limits on $\sigma(e^- \mu^+ \to e^- e^+ e^- \mu^+)$ are displayed in Fig.~\ref{fig:ul_gz}~(a) for null systematic uncertainty and in Fig.~\ref{fig:ul_gz}~(b) for a systematic uncertainty of $5\%$. For ${\cal L}=1000~{\rm fb}^{-1}$, the excluded cross-section ranges from $0.027$~fb at $m_{Z'}=5.5$~GeV to $0.13$~fb at $m_{Z'}=90$~GeV without systematics, and from $0.029$~fb to $0.18$~fb once a $5\%$ uncertainty is included. The shape of the curve closely follows that obtained for the ALP, with a flat behaviour at low mass, a local worsening at $m_{Z'}=90$~GeV driven by the $Z\to e^+e^-$ background, and a marked improvement at $m_{Z'}=300$~GeV, where the limit returns to $0.027$~fb. The numerical values are also close to the ALP ones, within $25\%$ over the whole range, the two effects discussed above compensating each other to a large extent: the lower signal efficiency of the $Z'$ selection is offset by the correspondingly smaller background admitted by the narrower signal window.

The corresponding limits on the $Z'$-electron coupling are shown in Fig.~\ref{fig:ul_gz}~(c) and (d). For ${\cal L}=1000~{\rm fb}^{-1}$ and in the absence of systematic uncertainty, the reach varies between $g_{Z'}\lesssim4.2\times10^{-5}$ and $1.8\times10^{-3}$ for $5.5~{\rm GeV}\leq m_{Z'}\leq100$~GeV, and between $3.5\times10^{-3}$ and $1.1\times10^{-2}$ for $150~{\rm GeV}\leq m_{Z'}\leq300$~GeV. Including a $5\%$ systematic uncertainty relaxes these values only mildly, to $4.4\times10^{-5}-2.2\times10^{-3}$ and $3.9\times10^{-3}-1.1\times10^{-2}$, respectively. As in the ALP case, the monotonic degradation of the coupling limit with increasing mass reflects the phase-space suppression of the production cross-section rather than any loss of analysis performance, the variation of the signal efficiency being comparatively minor.

\section{Interpretation of the results}
\label{sec:interpretation}

\begin{figure}[tb!]
    \centering 
    \includegraphics[scale=0.3]{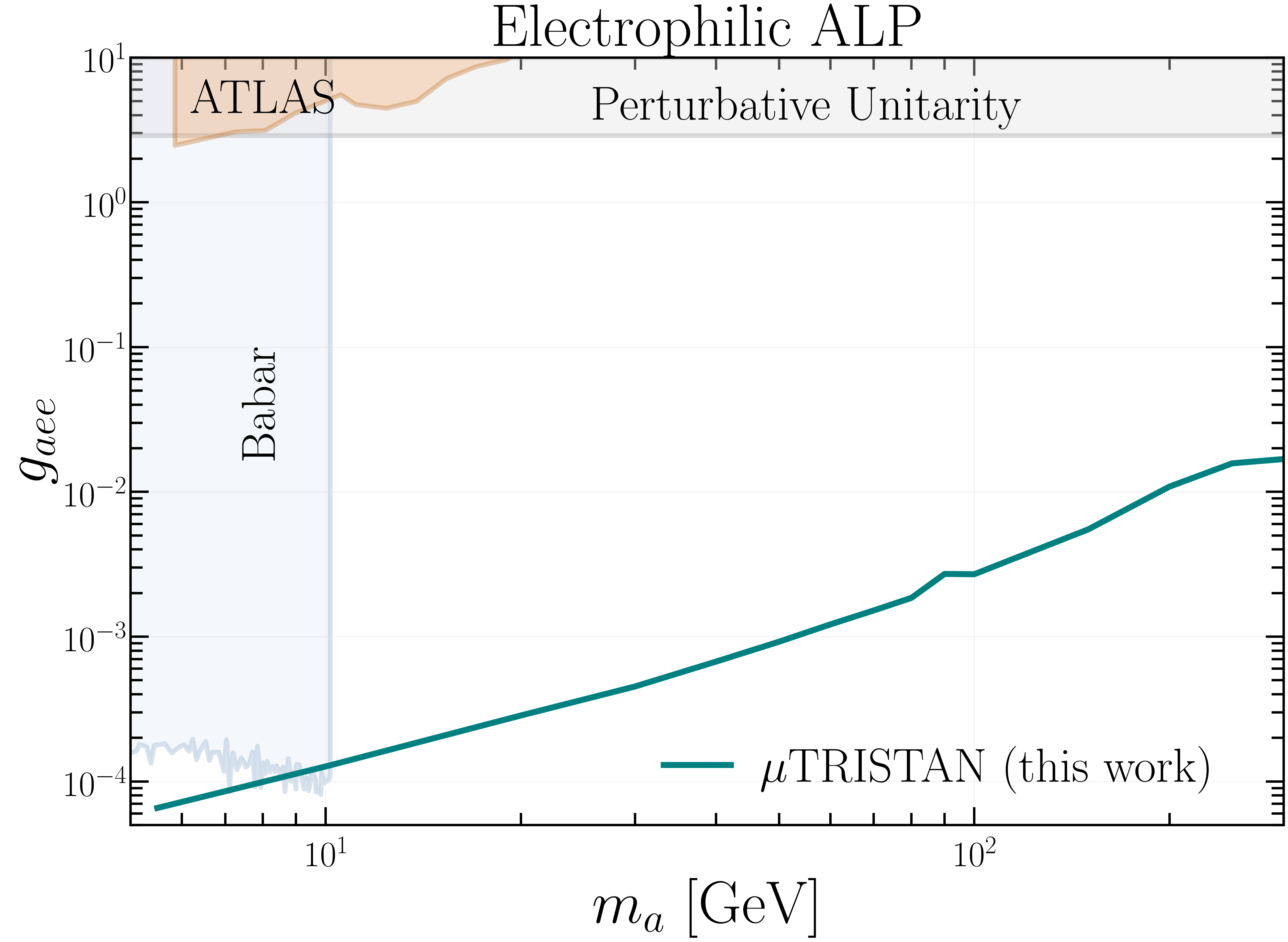}
    \caption{The projected $95\%$ CL upper limit on $g_{aee}$, without any systematics and $\mathcal{L}=1000~\mathrm{fb}^{-1}$, is shown as a function of the ALP mass, in teal color. Other existing constraints, such as the perturbative limit \cite{Cornella:2019uxs} and reinterpreted BaBar \cite{BaBar:2014zli} limit for the electrophilic mediator, are also shown. The constraint coming from ATLAS \cite{Biekotter:2025fll} through the loop-induced $g_{a\gamma\gamma}$ coupling is also shown.}
    \label{fig:com_ge}
\end{figure}

\begin{figure}[tb!]
    \centering 
    \includegraphics[scale=0.3]{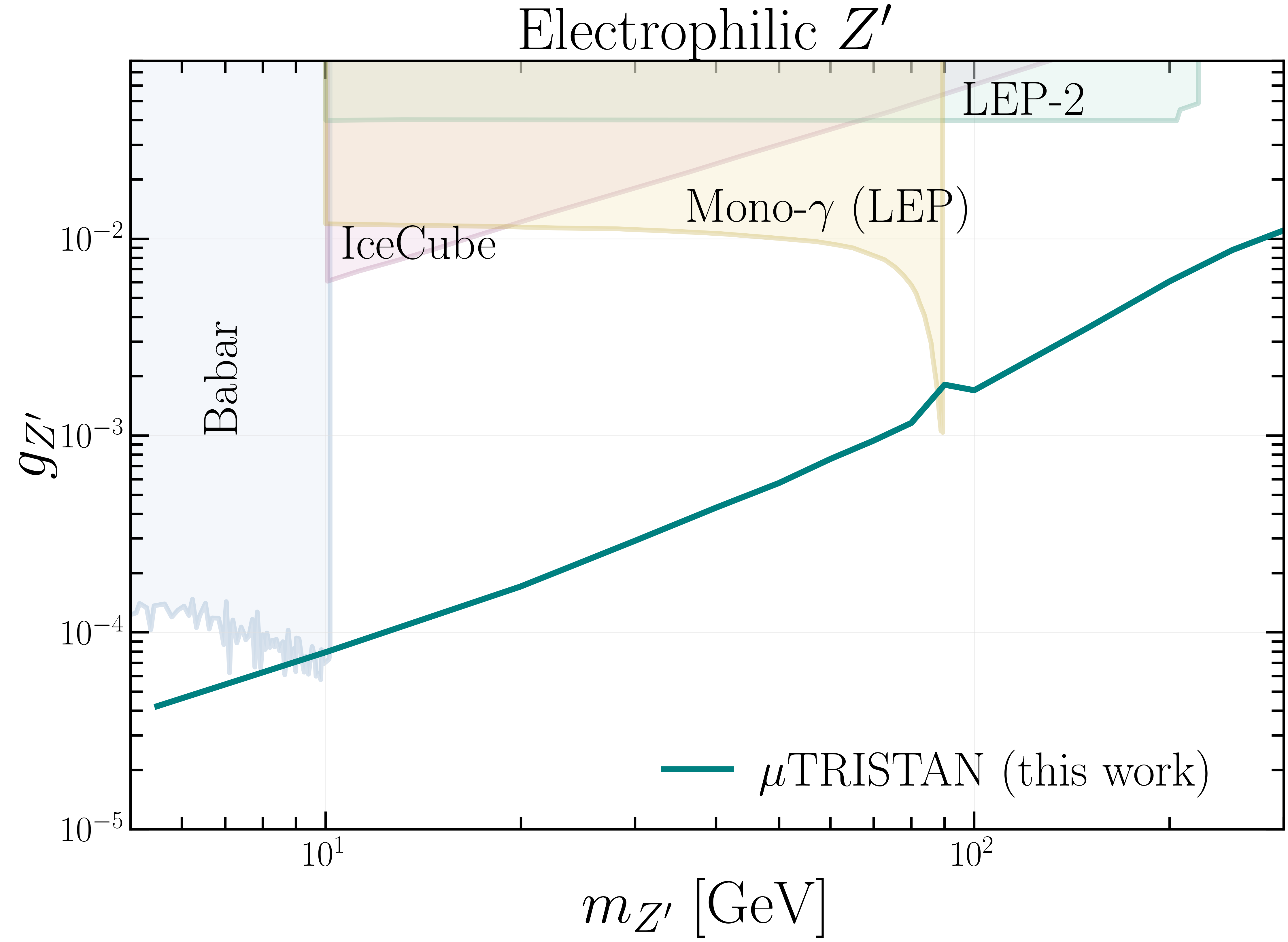}
    \caption{The projected $95\%$ CL upper limit on $g_{\rm Z'}$, without any systematics and $\mathcal{L}=1000~\mathrm{fb}^{-1}$, is shown as a function of the $Z'$ mass, with teal color. Other existing constraints like BaBar \cite{BaBar:2014zli}, LEP \cite{DELPHI:2003dlq,DELPHI:2008uka} are also displayed. Note that the IceCube constraint depends on the $\nu_e$ NSI \cite{IceCube:2022pbe}.}
    \label{fig:com_gz}
\end{figure}

In this section, we compare the projected sensitivities obtained in Sections~\ref{sec:collider-alp} and \ref{sec:analysis_zpr} with the constraints already available on electrophilic mediators. The comparison is presented in Fig.~\ref{fig:com_ge} for the ALP scenario and in Fig.~\ref{fig:com_gz} for the $Z'$ scenario. In both figures, the teal curve corresponds to the $95\%$ CL upper limit of Fig.~\ref{fig:ul_gaee_alp}~(c) and Fig.~\ref{fig:ul_gz}~(c), respectively, obtained for the highest integrated luminosity considered in this work, ${\cal L}=1000~{\rm fb}^{-1}$, and in the absence of systematic uncertainty on the background. The shaded regions indicate the parameter space already excluded by existing measurements.

We begin with the ALP scenario shown in Fig.~\ref{fig:com_ge}. Below the kinematic reach of a $B$~factory, $m_a\lesssim10$~GeV, the BaBar search for a light resonance decaying into an $e^+e^-$ pair~\cite{BaBar:2014zli}, excludes $g_{aee}\gtrsim(0.8$--$2)\times10^{-4}$. The projected $\mu$TRISTAN limit, $g_{aee}\lesssim6.5\times10^{-5}$ at $m_a=5.5$~GeV, is stronger by a factor of about $2.7$ at the lower end of this window. The limits become comparable above $m_a\simeq 8$ GeV. The BaBar exclusion does not extend to larger masses, since a $B$~factory operating at $\sqrt{s}\simeq10.6$~GeV cannot produce a heavier state.

Above the BaBar kinematic threshold, however, the available constraints are considerably less restrictive. The requirement of perturbative unitarity~\cite{Cornella:2019uxs} limits the coupling to $g_{aee}\lesssim3$, while the ATLAS constraint~\cite{Biekotter:2025fll}, which probes $g_{aee}$ only indirectly through the electron loop that generates $g_{a\gamma\gamma}$, becomes relevant for $g_{aee}\gtrsim2.5$ and therefore lies close to the perturbativity boundary. This behaviour can be understood from the discussion in Section~\ref{sec:theory}: the loop-induced di-photon coupling is proportional to the electron mass and is suppressed as $m_e/m_a^2$, so that searches based on photon final states are expected to retain little sensitivity to a purely electrophilic ALP in the mass range considered here. In contrast, the $e^-e^+e^-\mu^+$ final state probes $g_{aee}$ at the level of the tree-level vertex. In the mass range $5.5~{\rm GeV}\leq m_a\leq100$~GeV, the projected limits vary between $g_{aee}\lesssim6.5\times10^{-5}$ and $2.7\times10^{-3}$, while for $150~{\rm GeV}\leq m_a\leq300$~GeV they lie between $5.5\times10^{-3}$ and $1.7\times10^{-2}$. With respect to the perturbativity bound, this corresponds to an improvement of more than four orders of magnitude at the lower end of the mass range and of more than two orders of magnitude at the upper end, in a region of parameter space that is presently unconstrained by direct searches.

The $Z'$ scenario, shown in Fig.~\ref{fig:com_gz}, is subject to a larger number of existing constraints. This can be traced back to the fact that a $Z'$ coupled exclusively to the electron is not compatible with anomaly cancellation. In anomaly-free realisations, such as those based on $L_e-L_{\mu/\tau}$, the gauge boson also couples to $\nu_e$ with the same strength, and consequently acquires an invisible decay mode and contributes to neutrino observables, neither of which is the case for the electrophilic ALP. Both effects are reflected in the figure. The mono-photon analyses at LEP~\cite{DELPHI:2003dlq,DELPHI:2008uka}, which search for $Z'\to\nu\bar\nu$ produced in association with a photon, exclude $g_{Z'}\gtrsim1.2\times10^{-2}$ at $m_{Z'}=10$~GeV. The bound remains nearly flat up to $m_{Z'}\simeq75$~GeV and then strengthens rapidly, reaching $g_{Z'}\simeq1\times10^{-3}$ close to the $Z$ mass, beyond which the single-photon searches lose sensitivity. The IceCube measurement of non-standard neutrino interactions~\cite{IceCube:2022pbe} constrains the same $\nu_e$ coupling; it is most restrictive at low mass, $g_{Z'}\gtrsim6\times10^{-3}$ around $m_{Z'}=10$~GeV, and weakens progressively for heavier mediators. Finally, the LEP-2 bound~\cite{DELPHI:2003dlq,DELPHI:2008uka} is essentially independent of the mass, $g_{Z'}\simeq4\times10^{-2}$, as expected for a constraint of contact-interaction type, and starts to weaken as $m_{Z'}$ approaches the LEP-2 centre-of-mass energy.

Compared with these bounds, the projected $\mu$TRISTAN sensitivity is stronger over essentially the whole mass range. Below $10$~GeV, the BaBar bound, $g_{Z'}\gtrsim(0.6$--$1.4)\times10^{-4}$, is weaker than the projected limit by a factor of about three at $m_{Z'}=5.5$~GeV. The two become comparable for $m_{Z'}\simeq 8$~GeV and above. At $m_{Z'}=10$~GeV, the limit $g_{Z'}\lesssim7.9\times10^{-5}$ improves upon the IceCube and LEP mono-photon constraints by factors of about $80$ and $150$, respectively, while at $m_{Z'}=50$~GeV the improvement over the mono-photon bound is close to a factor of $20$. At $m_{Z'}=100$ and $200$~GeV, the projected limits are stronger than the LEP-2 bound by factors of approximately $25$ and $7$, respectively, and above $m_{Z'}\simeq220$~GeV, where the LEP-2 exclusion terminates, the present analysis provides the only constraint on this parameter space. An exception occurs in the narrow dip of the mono-photon exclusion at $m_{Z'}\simeq89$~GeV, where the LEP bound reaches $g_{Z'}\simeq1.0\times10^{-3}$ and remains slightly stronger than the projected value of $1.8\times10^{-3}$; this is also the mass at which our own sensitivity is locally degraded by the $Z\to e^+e^-$ component of the background entering the signal window. We note, however, that the two constraints are of a different nature, since the mono-photon bound applies to the invisible decay mode and therefore to the neutrino coupling, whereas the $e^-e^+e^-\mu^+$ final state probes the visible $Z'\to e^+e^-$ vertex directly. More generally, the mono-photon and IceCube regions displayed in Fig.~\ref{fig:com_gz} are obtained under the assumption of an unsuppressed $\nu_e$ coupling, while the limits derived in this work do not rely on it.

Comparing the two scenarios is instructive, because the two analyses share the same final state, the same reconstruction chain and the same statistical treatment. At the level of the production cross-section, the two limits are nearly the same, agreeing to within $25\%$ over the entire mass range, as can be seen by comparing Fig.~\ref{fig:ul_gaee_alp}~(a) with Fig.~\ref{fig:ul_gz}~(a). This is not accidental: the two effects that distinguish the scenarios at the level of the event selection act in opposite directions and largely cancel, the lower pairing efficiency of the $Z'$ reducing the signal efficiency while its narrower resonance admits a correspondingly smaller background within the $\pm1\sigma$ window. The reach in the coupling, on the contrary, is systematically better for the vector, $g_{aee}/g_{Z'}=1.5-1.8$ across the benchmarks. Since the cross-section limits are almost equal, this difference is entirely a property of the production mechanism rather than of the analysis: at a fixed coupling the vector is produced with a cross-section about $2.3$ times larger than the pseudoscalar, and because $\sigma\propto g^{2}$ the coupling limit improves by roughly the square root of that ratio.

\section{Conclusion}
\label{sec:conclusion}

In this work, we have investigated the sensitivity of the proposed $\mu$TRISTAN collider to electrophilic new bosons in the mass range between $5.5$ and $300$~GeV. Within an effective field theory framework, we consider two representative scenarios: a pseudoscalar axion-like particle (ALP $(a))$ coupled to the electron via a Yukawa-type interaction, and a $U(1)_X$ neutral vector boson $Z'$ coupled to the electron current. In both cases, we have studied the $e^-\mu^+\to e^-e^+e^-\mu^+$ final state at $\sqrt{s}=346$~GeV, corresponding to beam energies of $E_e=30$~GeV and $E_\mu=1000$~GeV. The signal and background samples have been generated at leading order with \texttt{MadGraph5\_aMC@NLO}, using a \texttt{UFO} model implemented in \texttt{FeynRules}, and detector effects have been emulated with a Gaussian momentum smearing tuned to the resolution of the ILD detector concept. In both scenarios, the analysis is based on the invariant mass of the electron--positron pair produced in the decay: the pair is identified with a boosted decision tree trained to tag the spectator electron using production kinematics only, the resulting $m_{ee}$ peak is fitted with a double-sided Crystal Ball function, and the signal region is defined by the interval $[\mu-\sigma,\,\mu+\sigma]$ obtained from the fit. In both cases, the exclusion limits have been obtained from the Asimov significance formula, for four integrated luminosity configurations and for null and $5\%$ systematic uncertainty on the background.

For the ALP scenario, the $95\%$ CL upper limits on the signal production cross-section at ${\cal L}=1000~{\rm fb}^{-1}$ range from $0.03$ to $0.12$~fb without systematics, and from $0.03$ to $0.19$~fb when a $5\%$ systematic uncertainty is included. Translated into the ALP-electron coupling, these correspond to $g_{aee}\lesssim(6.5\times10^{-5}-2.7\times10^{-3})$ for $5.5~{\rm GeV}\leq m_a\leq100$~GeV, degrading to $g_{aee}\lesssim(5.5\times10^{-3}-1.7\times10^{-2})$ for $150~{\rm GeV}\leq m_a\leq300$~GeV. The reconstructed mass peak is well described by the Crystal Ball parameterisation, with a fitted position reproducing the generated mass to better than $0.15\%$ and a relative width of $\sigma/\mu=1.64-1.71\%$ that remains stable over the whole mass range. For the $Z'$ scenario, analysed with the same resonance-fit strategy, the corresponding cross-section limits vary between $0.027$ and $0.13$~fb without systematics, and the coupling reach is $g_{Z'}\lesssim(4.2\times10^{-5}-1.8\times10^{-3})$ for $5.5~{\rm GeV}\leq m_{Z'}\leq100$~GeV, degrading to $g_{Z'}\lesssim(3.5\times10^{-3}-1.1\times10^{-2})$ for $150~{\rm GeV}\leq m_{Z'}\leq300$~GeV. The reconstructed $Z'$ peak is narrower than the ALP one, $\sigma/\mu=1.09-1.24\%$ against $1.64-1.71\%$. In both scenarios, the degradation of the reach with increasing mediator mass is driven primarily by the reduction of the signal production cross-section, which becomes increasingly phase-space suppressed as the resonance mass approaches the kinematic limit of a $346$~GeV machine. The selection efficiency, which varies between $0.33$ and $0.45$ for the ALP and between $0.25$ and $0.43$ for the $Z'$, and whose mass dependence follows that of the pairing efficiency rather than of the production rate, plays a subleading role. The two scenarios yield very similar limits on the production cross-section, since the kinematics of the $e^-e^+e^-\mu^+$ final state are governed primarily by the mass of the intermediate resonance rather than by its spin, and since the lower pairing efficiency of the $Z'$ and the smaller background compensate each other to a large extent. The reach in the coupling is nevertheless better for the vector by a factor of $1.5-1.8$, which reflects its larger production cross-section at a fixed coupling rather than any difference in the performance of the analysis.

Comparing these projections with the existing constraints, we find that in the low-mass region, $m\lesssim10$~GeV, the projected sensitivity is comparable to that of the BaBar search, reinterpreted here for the electrophilic mediators, whose reach does not extend to heavier states because of the centre-of-mass energy of a $B$~factory. Above this threshold, the situation differs markedly between the two scenarios. For the electrophilic ALP, the only available bounds are the perturbative unitarity requirement and the ATLAS constraint acting through the loop-induced $g_{a\gamma\gamma}$ coupling, both of which are restrictive only for $g_{aee}\gtrsim2.5$; the projections obtained here therefore probe a region of parameter space that is presently unconstrained by direct searches, improving on the perturbativity bound by more than four orders of magnitude at low mass and by more than two orders of magnitude at the upper end of the range. For the $Z'$, whose anomaly-free realisations necessarily involve a coupling to $\nu_e$, the mono-photon searches at LEP, the LEP-2 contact-interaction bound and the IceCube measurement of non-standard neutrino interactions provide additional constraints; the projected $\mu$TRISTAN sensitivity nevertheless improves upon them by up to two orders of magnitude over most of the mass range, and provides the only constraint above $m_{Z'}\simeq220$~GeV, where the LEP-2 exclusion terminates.

A few remarks are in order regarding the limitations of the present study. The analysis is performed at the parton level, with detector effects emulated using ILC-like resolutions; a more realistic treatment would require a dedicated $\mu$TRISTAN detector design, together with an assessment of the beam-induced backgrounds associated with a stored muon beam. A further limitation is intrinsic to the reconstruction of the resonance: the assignment of the decay pair is correct in $83\%$ of the events on average for the ALP and $75\%$ for the $Z'$, with a minimum near $m\simeq150$~GeV, and the residual mis-pairing populates a broad, non-resonant component underneath the peak. Since this component is smooth and is largely removed by the narrow mass window, it does not bias the extraction of $\mu$ and $\sigma$, but a dedicated treatment of the mis-paired events, for instance through a same-sign $e^-e^-$ control sample, would allow it to be constrained directly from data. In addition, the comparisons shown in Figs.~\ref{fig:com_ge} and \ref{fig:com_gz} correspond to the case of null systematic uncertainty. Adopting instead the $5\%$ systematics configuration of Fig.~\ref{fig:ul_gaee_alp}~(d) and Fig.~\ref{fig:ul_gz}~(d) weakens the coupling limits by at most about $25\%$, the effect being largest at $m\simeq90$ and $200$~GeV, where the background surviving the mass window is largest, and almost negligible at $m=300$~GeV, where it is smallest; the overall picture therefore remains qualitatively unchanged. It should also be kept in mind that the external constraints shown here have been obtained from analyses with different statistical treatments and underlying assumptions, and that the comparison is intended to be indicative rather than the result of a combined fit. With these caveats, the results presented above suggest that the mass window between the kinematic reach of the $B$~factories and the contact-interaction limits from LEP, which is currently the least constrained region for electrophilic mediators, can be probed effectively at an asymmetric $\mu^+e^-$ collider with a clean leptonic final state.

\section*{Acknowledgement}
The work of AA is funded by the Indian Association for the Cultivation of Science, Kolkata.

\bibliographystyle{JHEP}
\bibliography{refs}

\end{document}